%% file: loss.tex
\documentclass[twocolumn,aps]{revtex4-1}
\usepackage{graphicx} 
\pdfoutput=1

\usepackage{graphicx} 
\usepackage{amsmath}
\usepackage{amssymb}
\usepackage[mathletters]{ucs}
\usepackage[utf8x]{inputenc}		    
\usepackage[T1]{fontenc}		    
\usepackage{calc}
\usepackage{array}
\usepackage{float}

\usepackage{xstring}
\usepackage{tikz}
\usepackage{xifthen}
\usepackage{etoolbox}
\usepackage{braket}
\usepackage{mathtools}
\DeclarePairedDelimiter{\floor}{\lfloor}{\rfloor}

\begin{document}

\title{Practical Topological Cluster State Quantum Computing Requires Loss Below 1\%}

\author{Adam C. Whiteside$^{1}$, Austin G. Fowler$^{2,1}$}

\affiliation{
    $^1$Centre for Quantum Computation and Communication Technology, \\
    School of Physics, The University of Melbourne, Victoria, 3010, Australia \\
    $^2$Department of Physics, University of California, Santa Barbara, California 93106, USA
}

\date{\today}

\begin{abstract} 
    The surface code cannot be used when qubits vanish during computation;
    instead, a variant known as the topological cluster state is necessary. It
    has a gate error threshold of $0.75\%$ and only requires nearest-neighbor
    interactions on a 2D array of qubits. Previous work on loss tolerance using
    this code only considered qubits vanishing during
    measurement~\citep{Barrett:2010aa}. We begin by also including qubit loss
    during two-qubit gates and initialization, and then additionally consider
    interaction errors that occur when neighbors attempt to entangle with a
    qubit that isn't there. In doing so, we show that even our best case
    scenario requires a loss rate below $1\%$ in order to avoid
    considerable space-time overhead.
\end{abstract}

\maketitle

\setlength\parfillskip{0.5\linewidth plus 0.1\linewidth minus 0.1\linewidth}

\section{Introduction}
\label{sec:intro}

Quantum error correction (QEC) codes have made it plausible to perform
arbitrarily robust quantum computation using imperfect hardware. Each gate must
be below a threshold error rate, which is unique to the code and types of error
present. At $\sim$$1\%$~\citep{Fowler:2012ac}, the \textit{surface
code}~\citep{Bravyi:1998aa} has the highest threshold of any code needing
only a 2D array of qubits with nearest-neighbor interactions -- the most
practical experimental requirements of any code with a similar threshold. 

Quantum hardware based on linear optics~\citep{Fortescue:2014aa}, optical
lattices~\citep{Vala:2005aa}, and
trapped-ions~\citep{Blakestad:2011aa,Wright:2013aa} suffers from
\textit{qubit loss}. Qubit loss -- or simply \textit{loss} -- is defined as
when a qubit vanishes during computation, but can be detected and replaced at
measurement. This differs from manufacturing faults, where a qubit is
permanently unusable, requiring alternate methods to work around.  Loss is
also different to \textit{leakage}, where a qubit transitions into a
non-computational state rather than disappearing entirely.

Unfortunately, the surface code cannot be used when loss can occur, regardless
of how unlikely. The presence of these errors necessitates the use of another
code, the \textit{topological cluster state}~\citep{Raussendorf:2003aa}. While
the topological cluster state is a 3D code, it can be implemented on a 2D
lattice with nearest neighbor interactions, just as with the surface code. It
therefore has the same physical requirements -- with a threshold of
$\sim$$0.75\%$~\citep{Fowler:2012aa} -- but can also continue to correct errors
when some qubits vanish. 

Initial work to determine how much loss is tolerable using the topological
cluster state was promising -- indicating that the loss threshold was
$\sim$$24.9\%$~\citep{Barrett:2010aa}. This threshold is true under a model
where loss occurs only during measurement. However, of the hardware
implementations where loss can occur, the qubits can become lost at any time
rather than only during
measurement~\citep{Wright:2013aa,Knill:2000aa,Mandel:2003aa,Olmschenk:2010aa}.

There has been limited analysis of what errors occur when two qubits are
intended to be interacted but one of the pair is
missing~\citep{Herrera-Marti:2010aa}. When a two-qubit entangling gate is used,
but one of the pair has vanished, it is unclear what occurs to the remaining
qubit. Depending on the implementation of the hardware, the qubits neighboring
a lost one may acquire additional errors during two-qubit interactions.

The results presented here are based on current experimental targets for
computational error, $p \in \{10^{-3}, 10^{-4}\}$. While the thresholds under
our models are greater than $1\%$, we consider the amount of loss to be
practical only if a quantum computer would require fewer than twice as many
qubits when compared to one without any loss at all. Further details are
provided in the Results (Sec.~\ref{sec:results}).

To make this paper self contained, background information has been included
covering stabilizers in Sec.~\ref{sec:st}, cluster states in Sec.~\ref{sec:cs},
the topological cluster state in Sec.~\ref{sec:tcs}, error correction in
Sec.~\ref{sec:ec}, and how loss is handled in Sec.~\ref{sec:hql}. Those
familiar with the topological cluster state are invited to skip to
Sec.~\ref{sec:models} where we describe the models for loss we have used.
Sec.~\ref{sec:overhead} outlines how we have calculated overhead.
Sec.~\ref{sec:sim} discusses how the simulations were performed,
Sec.~\ref{sec:results} includes our results, and Sec.~\ref{sec:discussion} the
discussion. 

\section{Stabilizers}
\label{sec:st}

The stabilizer~\citep{Gottesman:1997aa} of a state $\ket{ψ}$ is the group
$\mathcal{G}$ of operators $A_i$, called stabilizers, that act on $\ket{ψ}$
without modifying it. 

\[
    A_i\ket{ψ} = \ket{ψ}
\]

Without the presence of errors, we know our state $\ket{ψ}$ is in the
simultaneous $+1$ eigenstate of the stabilizers. This, as will be explained,
allows a generating set for the group $\mathcal{G}$ of stabilizers to be used
to represent the state. For example, the $\hat{Z}$ operator is a stabilizer
for the state $\ket{0}$ and a generator of its stabilizer group.  This implies
that if we know that our state is stabilized by the $\hat{Z}$ operator, then
our state is the $+1$ eigenstate of $\hat{Z}$, namely $\ket{0}$. We will only work
with stabilizers that are tensor products of the $\hat{X}$ (bit flip),
$\hat{Z}$ (phase flip), $\hat{Y}$ (bit and phase flip) and $\hat{I}$ (identity)
operators.

In what is a somewhat confusing use of terminology, \textit{the stabilizers} of a
state $\ket{ψ}$ often refers instead to $\mathcal{S}$, a generating set of
$\mathcal{G}$. This \textit{stabilizer generator} set will typically have $n$ elements,
where $n$ is the number of qubits in the state $\ket{ψ}$.

In line with this, \textit{a stablizer} of the state
$\ket{ψ}$ often refers to an element $A$ of the stabilizer generator,
$\mathcal{S}$ ($A \in \mathcal{S}$). By definition, the element $A$ is also an
element of $\mathcal{G}$ ($A \in \mathcal{G}$). This terminology is the
result of more commonly working with the generating set $\mathcal{S}$ and the
elements of $\mathcal{S}$ rather than with the entire group.

There are many subsets of $\mathcal{G}$ that form a valid generating set. Due
to the closure of the group $\mathcal{G}$, any valid generating set
$\mathcal{S}$ can be transformed into another generating set $\mathcal{S}'$
by repeatedly taking the matrix product of elements of $\mathcal{S}$ and
replacing one of these elements with the result. Henceforth, the one chosen
is the most convenient set to work with. 

Stabilizers are of interest because they allow us to represent some complex
entangled states without having to write the actual state. The stabilizer
formalism cannot be used to express all quantum states, but it is sufficient
when defining the topological cluster state. In order to see how we can
represent a state $\ket{ψ}$ by its stabilizers, we first need to see how the
stabilizing set evolves when a unitary operator $U$ is applied.

\begin{align}
    A \ket{ψ} &= \ket{ψ} \\
    UA \ket{ψ} &= U \ket{ψ}  \\ 
    UAU^{†}U \ket{ψ} &= UU^†U \ket{ψ} \\
    (UAU^{†})U \ket{ψ} &= U \ket{ψ}
\end{align}

As $U$ is unitary, $U^{†}U$ is equal to the identity operator, and hence can be
introduced without modifying the equation. The result of this shows that if $A$
stabilizes $\ket{ψ}$, then $UAU^†$ stabilizes $U\ket{ψ}$. Therefore, rather
than looking at how the state $\ket{ψ}$ evolves, we can instead find a
stabilizing set for $\ket{ψ}$ and track how the stabilizers
evolve~\citep{Aaronson:2004aa,Anders:2006aa}.

Consequently, we can represent a state using $n$ stabilizers, each of $n$
operators, requiring only $n^2$ memory to store the state $\ket{ψ}$. This is
superior to potentially having to store the amplitudes for $2^n$ states if one
was to track the evolution of $\ket{ψ}$ itself.

The most basic examples of stabilizers are the $\hat{X}$ operator on the $\ket{+}$
state, $\hat{X}\ket{+} = \ket{+}$, and $\hat{Z}$ operator on the $\ket{0}$ state,
$\hat{Z}\ket{0}
= \ket{0}$. A basic 2 qubit example is the state
$\frac{\ket{00}+\ket{11}}{\sqrt{2}}$ that can be represented by the
stabilizers:

\begin{center}
    \renewcommand{\arraystretch}{1.1}
    \begin{tabular}{ >{$}l<{$} | >{$}c<{$} >{$}c<{$} >{$}c<{$} }
      A_1 & X_{1} & \otimes & X_{2} \\
      A_2 & Z_{1} & \otimes & Z_{2} \\
  \end{tabular}
\end{center}

\section{Cluster States}
\label{sec:cs}

A cluster state is any where each qubit is initialized in the $\ket{+}$ state,
and then at least one $C_Z$ gate performed. We will use the stabilizer
formalism to describe our cluster states. There is a convenient
way to determine a set of generators for a given \textit{cluster state}
$\ket{ψ}$:

\[
    A_i = X_i \otimes_{q_j\in\text{nghb}(q_i)}Z_j,
\]

\noindent where $i$ indexes a qubit in the state $\ket{ψ}$ and $\text{nghb}(q_i)$
is the set of qubits connected to $q_i$ via $C_Z$ gates. 

Consider the 2D cluster state in Fig.~\ref{fig:facestabilizer}. Each qubit
is initialized to the $\ket{+}$ state, which is stabilized by $\hat{X}$. Since each
qubit is currently independent, we can stabilize the 5-qubit state $\ket{ψ} =
\ket{+}_{1}\ket{+}_{2}\ket{+}_{3}\ket{+}_{4}\ket{+}_{5}$ with the following stabilizers:

\begin{figure}
\begin{center}
\resizebox{40mm}{!}{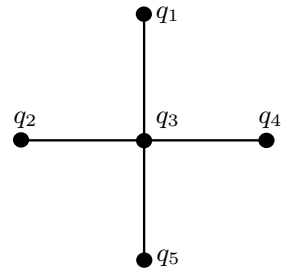}
\end{center}
\caption{
    The 5 qubit cluster state found on each of the faces of a 3D
    topological cluster state cell (Fig.~\ref{fig:tcscell}). Black dots are the
    qubits $q_1$ to $q_5$, initialized in the $\ket{+}$ state. Black lines
    indicate $C_Z$ gates between the two connected qubits.    
}\label{fig:facestabilizer}
\end{figure}

\begin{center}
    \renewcommand{\arraystretch}{1.1}
    \begin{tabular}{ >{$}l<{$} | >{$}c<{$} >{$}c<{$} >{$}c<{$} >{$}c<{$} >{$}c<{$} >{$}c<{$} >{$}c<{$} >{$}c<{$} >{$}c<{$} }
      A_1 & X_{1} & \otimes & I_{2} & \otimes & I_{3} & \otimes & I_{4} & \otimes & I_{5} \\
      A_2 & I_{1} & \otimes & X_{2} & \otimes & I_{3} & \otimes & I_{4} & \otimes & I_{5} \\
      A_3 & I_{1} & \otimes & I_{2} & \otimes & X_{3} & \otimes & I_{4} & \otimes & I_{5} \\
      A_4 & I_{1} & \otimes & I_{2} & \otimes & I_{3} & \otimes & X_{4} & \otimes & I_{5} \\
      A_5 & I_{1} & \otimes & I_{2} & \otimes & I_{3} & \otimes & I_{4} & \otimes & X_{5}
  \end{tabular}
\end{center}

\noindent Then, we can see how the application of $C_Z$ affects the stabilizers on two
qubits:

\begin{align}
    C_Z &= C_Z^{\dagger} \\
    \label{eq:czprop:1}
    C_Z(I \otimes X)C_Z &= Z \otimes X \\
    \label{eq:czprop:2}
    C_Z(X \otimes I)C_Z &= X \otimes Z \\
    \label{eq:czprop:3}
    C_Z(I \otimes Z)C_Z &= I \otimes Z \\
    \label{eq:czprop:4}
    C_Z(Z \otimes I)C_Z &= Z \otimes I
\end{align}

Following the $C_Z$ interactions outlined in Fig.~\ref{fig:facestabilizer}, we
want to take the stabilizers from state $\ket{ψ}$ and update them to stabilize
the state $\ket{ψ'} =
\Lambda_{1,3}\Lambda_{2,3}\Lambda_{4,3}\Lambda_{5,3}\ket{ψ}$ where
$\Lambda_{i,j}$ is a $C_Z$ applied between qubits $q_i$ and $q_j$. Based on the
definition of stabilizers, we know that if $A$ is a stabilizer of $\ket{ψ}$,
then $UAU^\dagger$ stabilizes $U\ket{ψ}$. Therefore,
$\Lambda_{1,3}A\Lambda_{1,3}^\dagger$ stabilizes $\Lambda_{1,3}\ket{ψ}$.
Repeating this for each of the $C_Z$ gates results in the following set of
stabilizers:

\begin{center}
    \renewcommand{\arraystretch}{1.1}
    \begin{tabular}{ >{$}l<{$} | >{$}c<{$} >{$}c<{$} >{$}c<{$} >{$}c<{$} >{$}c<{$} >{$}c<{$} >{$}c<{$} >{$}c<{$} >{$}c<{$} }
      A_1 & X_{1} & \otimes & I_{2} & \otimes & Z_{3} & \otimes & I_{4} & \otimes & I_{5} \\
      A_2 & I_{1} & \otimes & X_{2} & \otimes & Z_{3} & \otimes & I_{4} & \otimes & I_{5} \\
      A_3 & Z_{1} & \otimes & Z_{2} & \otimes & X_{3} & \otimes & Z_{4} & \otimes & Z_{5} \\
      A_4 & I_{1} & \otimes & I_{2} & \otimes & Z_{3} & \otimes & X_{4} & \otimes & I_{5} \\
      A_5 & I_{1} & \otimes & I_{2} & \otimes & Z_{3} & \otimes & I_{4} & \otimes & X_{5}
  \end{tabular}
\end{center}

Henceforth the tensor product symbol will be omitted for brevity. 

\section{Topological Cluster State}
\label{sec:tcs}

\begin{figure}[h]
\begin{center}
\resizebox{50mm}{!}{\includegraphics{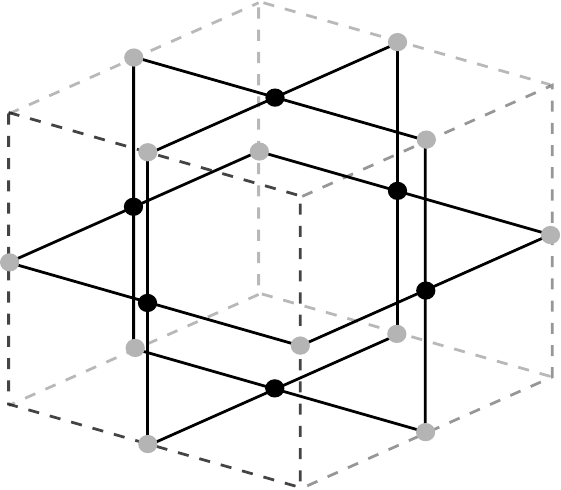}}
\end{center}
\caption{
    A Topological Cell State. The black and gray dots indicate qubits
    initialized in the $\ket{+}$ state, while solid lines indicate the
    application of a $C_Z$ gate. The black dots at the center of the faces are
    measured in the $\hat{X}$ basis ($M_X$) and form the measurement product
    of the cell. The dotted lines are a visual aid only.
}\label{fig:tcscell}
\end{figure}

The topological cluster state is composed of a 3D tiling of \textit{cells},
illustrated in Fig.~\ref{fig:tcscell}. Despite being a 3D tiling, the state can
be implemented on a 2D architecture with only nearest-neighbor interactions
(Fig.~\ref{fig:2Dimplementation}). A
single cell touches 18 qubits, each prepared in the $\ket{+}$ state, with $C_Z$
gates applied from each face qubit to each neighboring edge qubit. In this
configuration, there exists a stabilizer consisting of only $\hat{X}$
measurements on the qubits at the center of each face. When tiled in 3D, each
face qubit is shared with its neighbor, while each edge qubit is shared by a
total of four cells. Those unfamiliar with stabilizers and matrix products may
find an explicit derivation of the cell state aids their understanding
(Appendix.~\ref{app:cell}).

\begin{figure}
\begin{center}
\resizebox{60mm}{!}{\includegraphics{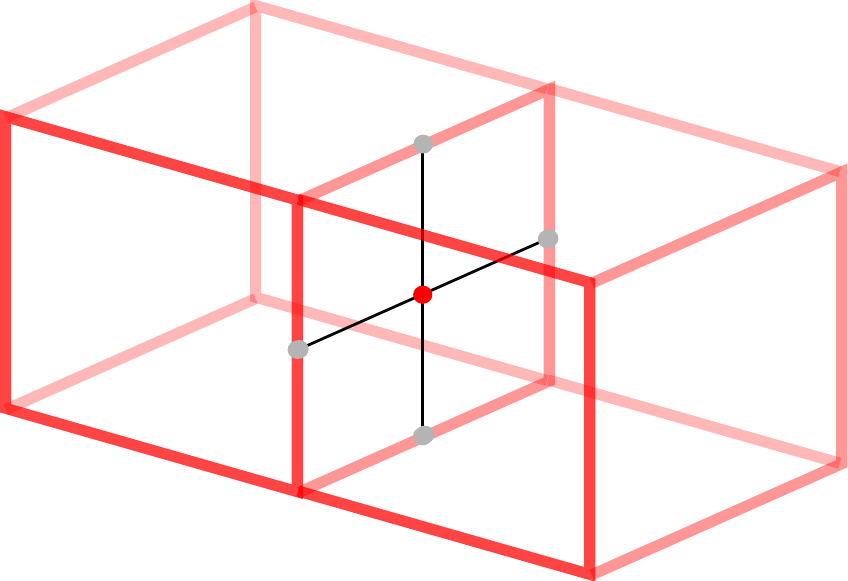}}
\end{center}
\caption{
    Detecting the location of an error using adjacent cells. The dark (red) dot is
    the qubit that has suffered an error. The face stabilizer shared by the
    cells has been shown, the others have been omitted for clarity. Each face
    qubit is shared by two adjacent cells. The thick lines indicate the
    boundaries of cells where a detection event occurs. The error causes a
    detection event to occur in both cells, allowing the location of the error
    to be inferred.
}\label{fig:twocell}
\end{figure}

When measuring the six face qubits in the $\hat{X}$ basis, each will either be
in the $+1$ or $-1$ eigenstate of $\hat{X}$. Due to the structure of the cell,
with no errors present, the cell will be in the $+1$ eigenstate of the
six face qubits. A $\hat{Z}$ error on one of the face qubits will flip the
eigenstate measured, causing the product of the six face measurements, called
the \textit{measurement product}, to become $-1$. Cells where an odd number of
errors have occurred can be identified by their $-1$ measurement products,
referred to as \textit{detection events}. As each face qubit is shared by a
neighbor, the location of errors can be inferred by the pattern of detection
events. If two neighboring cells are both measured to be in the $-1$ eigenstate
of the six face qubits, then it is likely that the qubit on the face shared by
the two neighbors has become erroneous (Fig.~\ref{fig:twocell}).

The topological cluster state only corrects $\hat{Z}$ and $M_{X}$ errors. This
is sufficient as $\hat{X}$ errors have no effect; an $\hat{X}$ error occurring
just before measurement does not change the measurement outcome, while one
occurring earlier will cause $\hat{Z}$ errors on one or more neighboring qubits
(Eqs.~\ref{eq:czprop:1}~--~\ref{eq:czprop:4}). Since each cell is stabilized by only
$\hat{X}$ basis measurements, we can ignore the $\hat{X}$ errors as long as we
correct the resultant $\hat{Z}$ errors~\citep{Fowler:2009ab}.

The full set of qubits tiled in both space and time is called a
\textit{lattice}. There exists a second lattice offset by one in both space
dimensions and in time, where each face qubit of the first lattice becomes an
edge qubit in the second, as well as the reverse. We arbitrarily call one the
\textit{primal lattice} and the other the \textit{dual lattice}. This allows
for complete error correction despite a single cell only detecting errors on
the face qubits; the eight dual cells that encompass a primal cell will have
the edge qubits as face qubits. An error is therefore classified as either
\textit{primal} or \textit{dual} depending on which lattice it is detected in.
These two interwoven lattices are also necessary for the implementation of two-qubit logical gates~\citep{Fowler:2009ab}. 

\begin{figure}
\begin{center}
\resizebox{60mm}{!}{\includegraphics{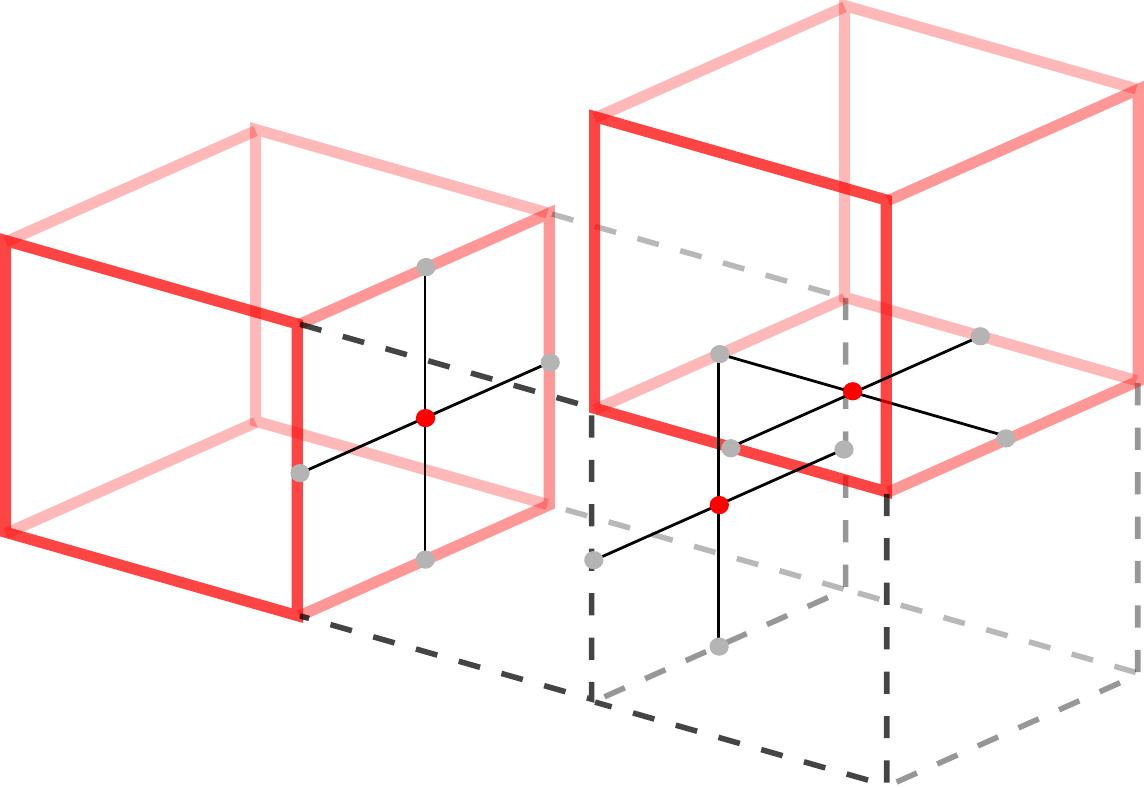}}
\end{center}
\caption{
    A chain of three errors, in space and time, affecting four cells. This
    demonstrates how chains can occur through multiple axes. The dark (red)
    dots indicate qubits where an error has occurred. Thick solid lines
    indicate the boundary of a cell where a detection event occurs. Thin 
    dotted lines indicate the boundary of a cell where no detection event
    occurs.
}\label{fig:fourcellL}
\end{figure}

The inference of errors from detection events is unfortunately inexact.
Multiple patterns of errors can create the same pattern of detection events due
to an even number of errors being undetectable to a cell. An even number of errors
will result in an even number of $-1$ measurements, leaving the cell in the
$+1$ eigenstate of the faces. Fig.~\ref{fig:fourcellL} shows a chain of
three errors, where only the cells at either end have a $-1$ measurement
product (and hence a detection event).

The calculation of detection events and the inference of errors will be done by
classical hardware running in parallel with a quantum
computer~\citep{Fowler:2012ac}. Given the observed pattern of detection events,
Edmonds' minimum weight perfect matching algorithm~\citep{Edmonds:1965ab} can be used
determine where errors are likely to have occurred. As the rate of
error increases, error combinations leading to the same detection events become
more frequent and incorrect inference becomes more common. It has been shown
that this algorithm can be parallelized to $\text{O}(1)$ given fixed computing
resources per unit area~\citep{Fowler:2013ac}, an essential property if the
classical software is to scale with the size of a large quantum computer. A
number of other methods are being studied in an attempt to get better
performance during
inference~\citep{Duclos-Cianci:2009aa,Duclos-Cianci:2014aa,Wootton:2012aa,Hutter:2014ab,Bombin:2012aa,Bravyi:2014ab}.

Logical qubits are formed in the topological cluster state by creating
\textit{defects}, where regions of the cluster state have their face
stabilizers measured in the $\hat{Z}$ basis. How logical qubits are formed and
manipulated to perform computation is unnecessary to understand the results
presented. We refer the reader to~\citep{Fowler:2009ab} for an overview, and to
the references~\citep{Raussendorf:2003aa,Raussendorf:2007ab,Raussendorf:2007aa}
for more details.

\section{Error Correction}
\label{sec:ec}

The \textit{distance} $d$ of the code is the minimum number of physical qubits
that need to be manipulated in order to connect two defects or encircle a
defect. If a chain of errors joins two defects of the same type (primal to
primal, or dual to dual), a logical error occurs. 

\begin{figure}
\begin{center}
\resizebox{40mm}{!}{\includegraphics{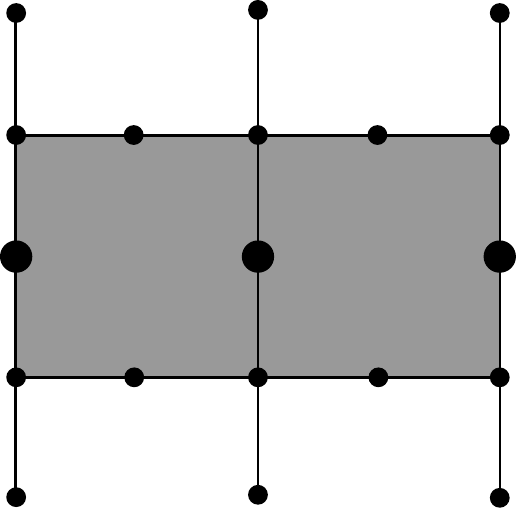}}
\end{center}
\caption{
    A single layer of a distance $3$ topological cluster state. Black dots
    indicate qubits, black lines indicate $C_Z$ gates. Gates between this layer
    and others are omitted for clarity. The gray squares are a visual indicator
    for the two complete dual cells. Dual boundaries are located to the left
    and right. The larger black dots are the three qubits where a chain of
    errors could connect the two dual boundaries. A second layer which is
    identical, except rotated 90 degrees, would have primal cells in gray and
    boundaries at the top and bottom.
}\label{fig:2d_lay1}
\end{figure}

A logical error also occurs if a defect is connected to a boundary, or if two
boundaries of the same type are connected. A \textit{boundary} is the smooth
edge of the lattice which ends in complete cells. In our chosen assignment of
primal and dual, the primal boundaries are at the top and bottom of the lattice
with the dual boundaries at the left and right (Fig~\ref{fig:2d_lay1}). 

Due to the inference in error correction, a logical error can occur with fewer
than $d$ errors. The minimum number of errors that can cause a logical error is
$d_e$:

\[
    d_e = \floor[\bigg]{\frac{d + 1}{2}}
\]

\begin{figure}
\begin{center}
\resizebox{60mm}{!}{\includegraphics{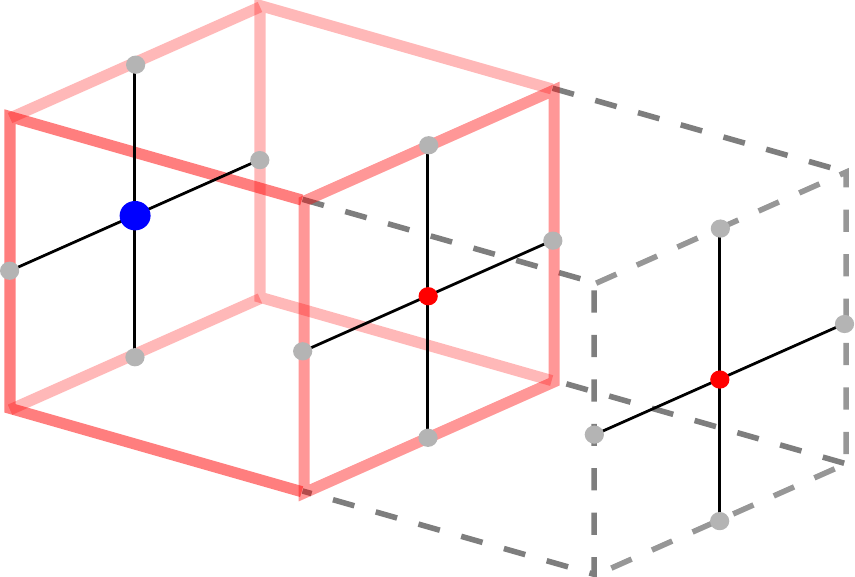}}
\end{center}
\caption{
    Two cells illustrating a distance $3$ fragment of a larger topological
    cluster state. The two ($d_e$) smaller dark dots are qubits with errors.
    The thick solid lines indicate the boundary of a cell with a detection
    event.  The larger dark dot indicates the most likely error pattern that
    would create the observed detection event. By applying a correction at the
    larger dark qubit, where one isn't needed, a chain spans the lattice
    resulting in a logical error.
}\label{fig:tcs_d3_logicalerror}
\end{figure}

Take the example of $d = 3$ as seen in Fig.~\ref{fig:tcs_d3_logicalerror};
$d_e$ in this instance is $2$. A line of $3$ qubits crosses the two cells. A
single error from one end has the same detection event pattern as two errors in
a row from the opposite end.  Therefore, when the two error case occurs with
probability $O(p^{d_e})$ the minimum weight perfect matching algorithm will
infer the single error case. By applying an erroneous correction, the algorithm
introduces another error completing the chain and connecting one edge of the
lattice to the other, forming a logical error.

For each set of errors that occupies at least $d_e$ positions along an axis,
there exists another set with equal or greater probability that will typically
be chosen by minimum weight matching, resulting in a logical error. In should
be emphasized that while the probability of a logical error $p_L$ is
proportional to $p^{d_e}$, the occurrence of $d_e$ errors does not always
result in a logical error.

\begin{figure}
\begin{center}
\resizebox{70mm}{!}{\includegraphics{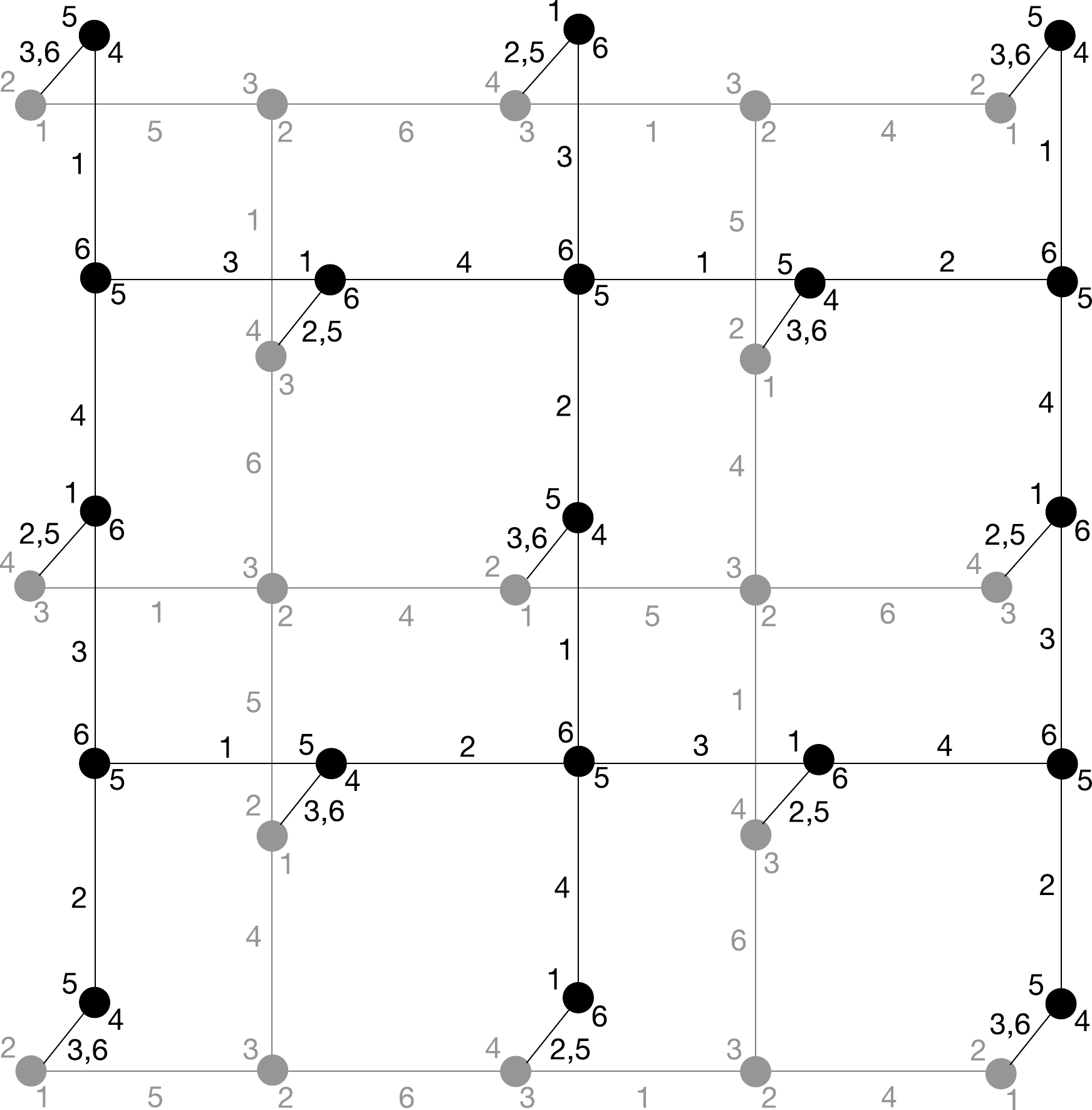}}
\end{center}
\caption{
    A distance $3$ topological cluster state on a 2D array of qubits.  Each
    line is a $C_Z$ gate and is labelled to indicate in which timestep(s) the gate
    is applied. The black dots indicate qubits. The number to the top left of a
    qubit indicates the timestep the qubit is initialized in, the number to the
    bottom right when it is measured.  Dark and light provide visual separation
    between the two layers. When implemented on a 2D architecture, no such
    distiniction exists, and the qubits are laid out physically as presented.
}\label{fig:2Dimplementation}
\end{figure}

By noting that each qubit is only interacted with its four neighbors, it is
possible to implement a 3D cell in only two layers
(Fig.~\ref{fig:2Dimplementation}). With careful timing of the $C_Z$ gates, each
qubit that needs to interact in the arbitrarily assigned up direction can be
initialized then interacted with its down neighbor. This neighbor will have then
completed all four of its interactions, allowing it to be measured and reinitialized.
This new qubit can then be interacted with the original qubit becoming its up
neighbor as well. This can be further implemented on a strictly 2D physical
architecture by interweaving the two layers into a single physical layer.

\section{Handling Qubit Loss}
\label{sec:hql}

\begin{figure}[htb]
\begin{center}
\resizebox{60mm}{!}{\includegraphics{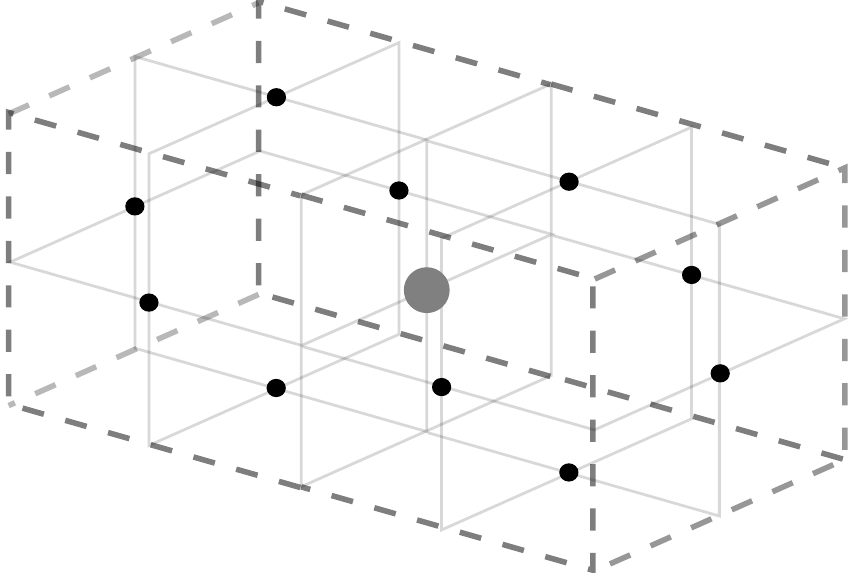}}
\end{center}
\caption{
    Two adjacent cells with a shared face qubit that has been lost. The dotted
    dark gray lines indicate the boundary of the merged stabilizer. The black
    dots indicate the ten qubits whose measurements will form the measurement
    product for the merged stabilizer. The large gray dot indicates the lost
    qubit.
}\label{fig:losstwocell}
\end{figure}

Lost qubits break the method of detecting errors outlined in Section
\ref{sec:tcs}. Consider again the case of two cells sharing a qubit that has
suffered an error (Fig.~\ref{fig:twocell}), however instead of an $\hat{X}$,
$\hat{Y}$, or $\hat{Z}$ error, the qubit was lost entirely. In this case, the
lost qubit cannot be measured, and neither cell would have six measurement
results. Given that both cells should be in the $+1$ eigenstate of all six
measurement results, the measurement products are meaningless without the last.

\begin{figure}[htb]
\begin{center}
\resizebox{60mm}{!}{\includegraphics{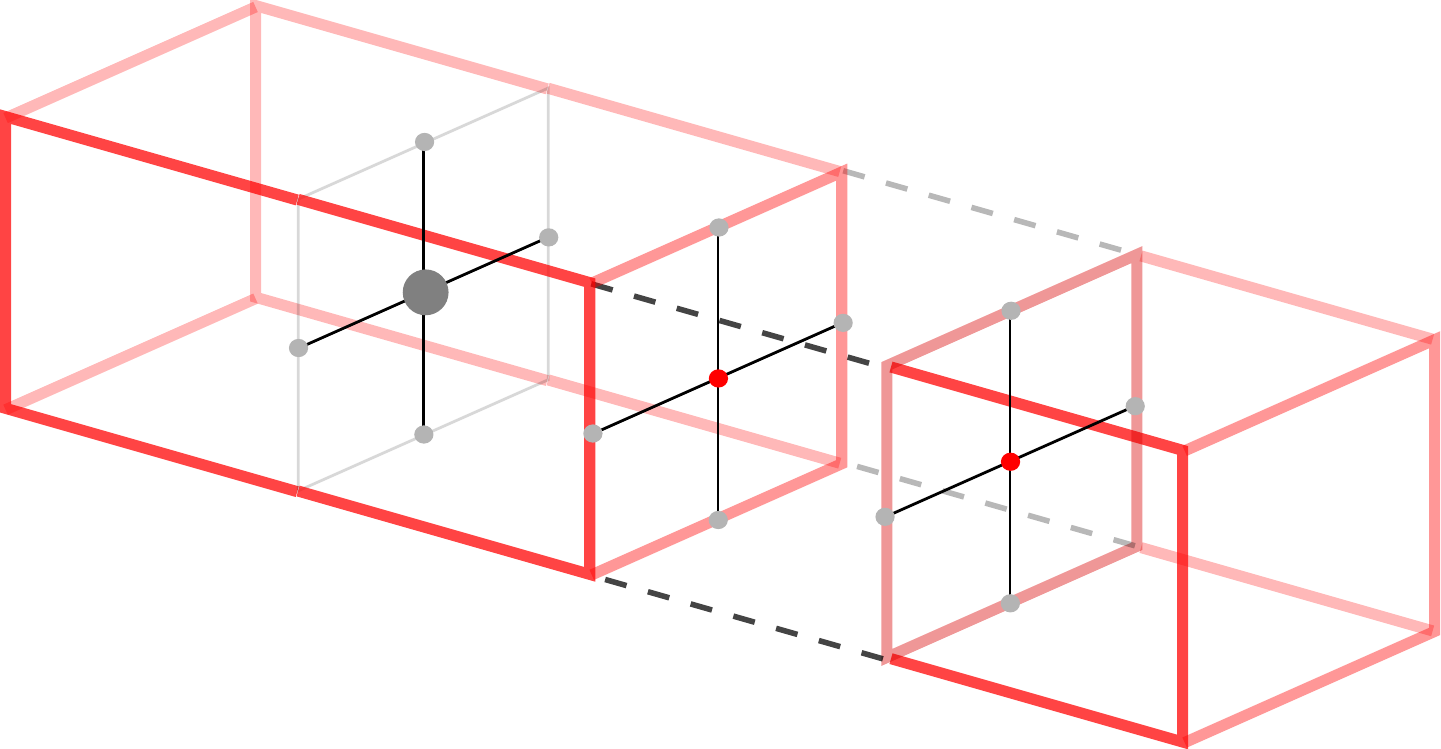}}
\end{center}
\caption{
    An error chain including a lost qubit error. The thick solid lines indicate
    cells where a detection event occurs. Small dark dots indicate errors. The large
    gray dot indicates the lost qubit. Note that the merged stabilizer due to the
    lost qubit behaves like a regular cell, with the detection events occurring
    at both ends of the regular error chain.
}\label{fig:lossfourcell}
\end{figure}

To allow for loss, we take advantage of the product of two stabilizers also
being a stabilizer. The product of two cell stabilizers forms a new stabilizer
consisting of the remaining ten face qubits, and is independent of the shared
qubit (Fig.~\ref{fig:losstwocell}). We cannot correct the loss error, however
this process can yield a valid detection event with which to detect errors on
the remaining qubits (Fig.~\ref{fig:lossfourcell}). Connected sets of loss
errors repeat this process creating larger and larger stabilizers. The
merging of stabilizers effectively reduces the number of computational errors
that can be tolerated.  This highlights the importance of the topological
cluster state, as its short error correction cycle followed by measuring and
replacing all qubits puts a limit on how long the reduced distance is in
effect.

\section{Modeling Loss}
\label{sec:models}

Each qubit in the topological cluster state undergoes the same gate sequence,
albeit staggered in time. This sequence includes initialization,
measurement, identity (storage), Hadamard, and controlled-Z ($C_Z$). 

In our models, and those we compare to, each gate introduces computational
errors with equal probability, $p_{comp}$. These errors manifest as unintended
$\hat{I}$, $\hat{X}$, $\hat{Y}$, or $\hat{Z}$ operations on the involved
qubit(s), each with an equal chance of occurring. This model of computational
error is identical to that published previously~\citep{Fowler:2012aa}.

Previous generic analysis of the topological cluster state and its tolerance to
loss only considered the performance when qubits were lost during
measurement~\citep{Barrett:2010aa}, though a hardware-specific analysis
has considered loss occurring during both initialization and
measurement~\citep{Herrera-Marti:2010aa}.

For our analysis loss is assumed to be detectable only at measurement and
occurs with equal probability $p_{loss}$ on all gates, except Hadamard, where
we assume there are no loss events. Single qubit Hadamard gates are typically
much simpler and faster than other gates, contributing negligible loss. With ion trap quantum hardware, ion movement and traversing
junctions are now well-controlled, introducing very little additional loss over
baseline background gas collisions. Since such collisions can occur at any
time, loss can occur during any gate, making loss less frequent during faster
gates~\citep{Wright:2013aa}.

When considering linear optical hardware, deterministic generation and
measurement of photons is extremely difficult, while two-qubit
gates are challenging as they require photon memory and feedforward processing. These processes
are therefore associated with significant photon loss. By contrast,
single-qubit gates require a simple beam splitter, with negligible
loss~\citep{Knill:2000aa}. This assumption holds true for optical lattice
hardware as well, where the single qubit rotation of the Hadamard gate is much
faster than the other gates~\citep{Mandel:2003aa,Olmschenk:2010aa}. 

\begin{figure}[tb]
\resizebox{80mm}{!}{\includegraphics{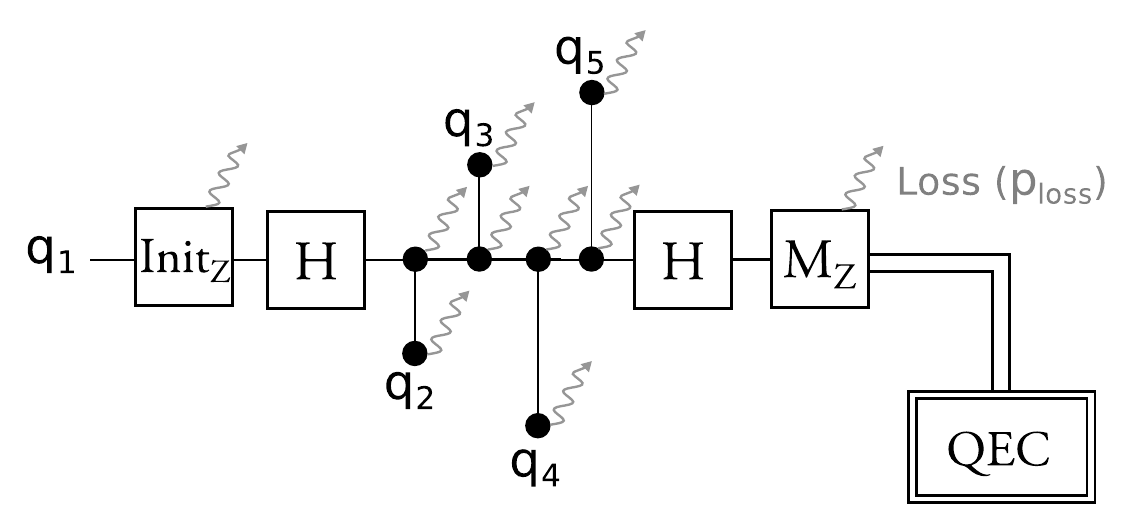}}
\caption{
    A round of error correction for a single qubit using the topological
    cluster state, with loss occurring at initialization, two-qubit gates, and
    measurement. The qubit $q_1$ is initialized in the $\hat{X}$ basis, interacted
    with its four neighbors using $C_Z$ gates, and then measured. The
    measurement result is passed to a classical computer for error correction.
    The gray arrows indicate where a qubit may be lost with probability
    $p_{loss}$.
}\label{fig:tcs_one_round_our_loss}
\end{figure}

In our models, loss events can occur after initialization, after each $C_Z$
gate and during measurement (Fig.~\ref{fig:tcs_one_round_our_loss}).  This
results in an average of $6 p_{loss}$ loss events per qubit, per round of error
correction. This is six times more than a model considering only loss during
measurement.

\begin{figure}[tb]
\resizebox{80mm}{!}{\includegraphics{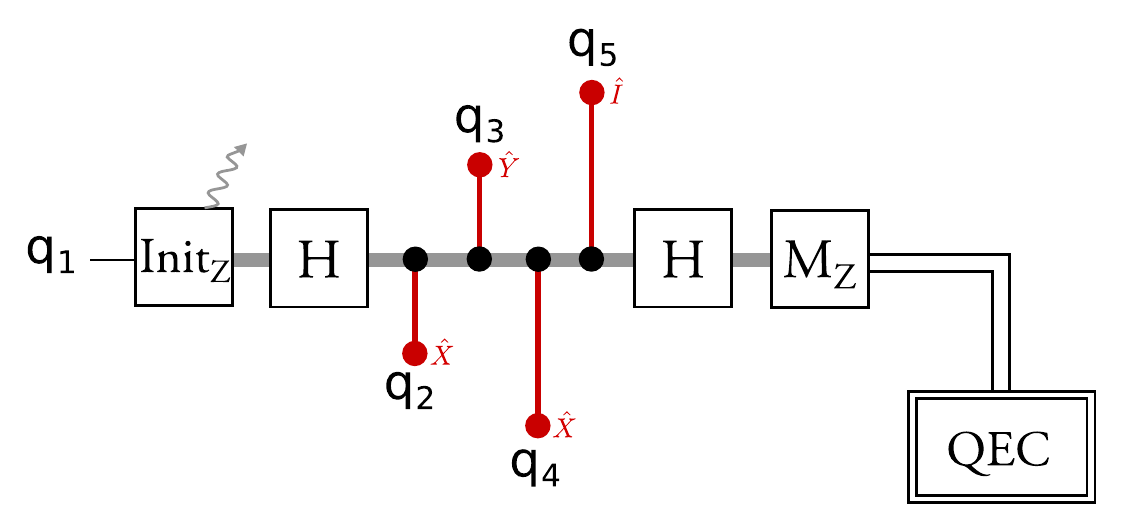}}
\caption{
    Demonstration of the effect of loss interaction errors. The grey arrow
    indicates where the qubit is lost. The thick grey line indicates the time line of the
    lost qubit through the circuit. With $p_{lint} = 1$, each neighbor ($q_2$,
    $q_3$, $q_4$, $q_5$) acquires an $\hat{I}$, $\hat{X}$, $\hat{Y}$, or
    $\hat{Z}$ error. In this instance $q_2$ and $q_4$ acquire an $\hat{X}$
    error, $q_3$ acquires $\hat{Y}$ and $q_5$ acquires $\hat{I}$.
}\label{fig:tcs_one_round_our_loss_interact}
\end{figure}

The second of our two error models also introduces loss interaction errors, which occur with
probability $p_{lint}$. A two-qubit gate, such as the $C_Z$ gate, assumes the
interaction between two present qubits. If one of the two qubits is missing,
the gate may fail and introduce error on the remaining qubit
(Fig.~\ref{fig:tcs_one_round_our_loss_interact}), we which refer to as a
\textit{loss interaction error}. We choose $p_{lint} = 1$, such that any qubit
being interacted with a lost qubit will always acquire an $\hat{I}$, $\hat{X}$,
$\hat{Y}$, or $\hat{Z}$ error, upper bounding the behaviour. Some hardware
implementations may be able to reduce, or entirely avoid this error by
constructing a gate that can only work in the presence of two qubits.

\section{Measuring Overhead}
\label{sec:overhead}

\begin{figure}[tb]
\resizebox{50mm}{!}{\includegraphics{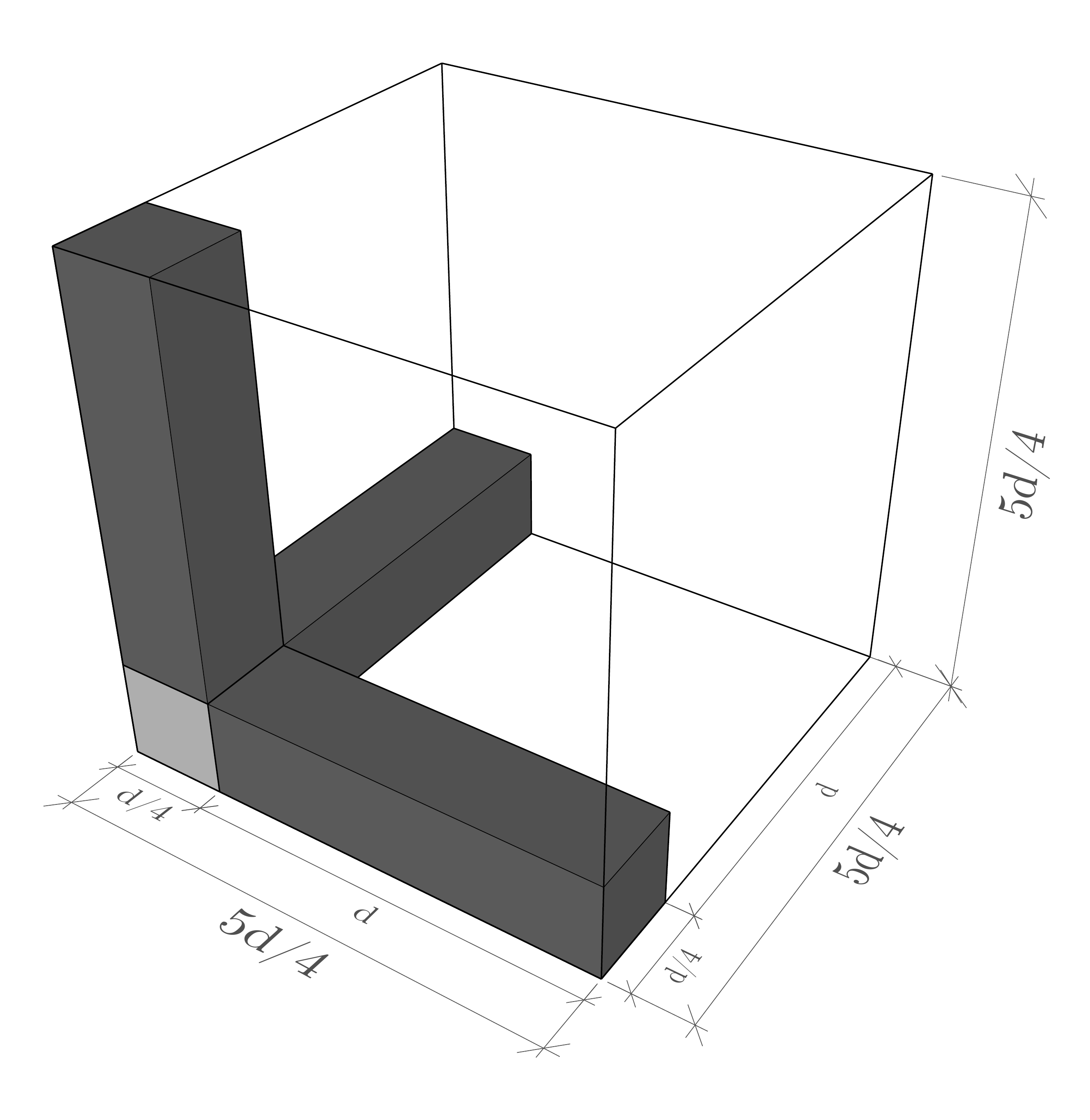}}
\caption{
    A plumbing piece with volume $(5d/4)^3$. This defect example contains a
    cubic fragment (light gray) of circumference $d$ (edge length $d/4$) and
    with prisms of length $d$ separating it from all neighbors (dark gray). 
}\label{fig:plumbing}
\end{figure}

In order to calculate the overhead of loss we define a volume of space-time
known as a \textit{plumbing piece}. Consider a square defect of circumference
$d$. Such a defect must be distance $d$ from all other defects in space and
time. A space-time volume containing this defect, and all necessary space
around it (see Fig.~\ref{fig:plumbing}) is a plumbing piece and will have an
edge length (in cells) of:

\[
    n \approx \frac{5d}{4}
\]

Each cell touches 18 qubits. Given that each face qubit is shared by its
neighbor, and each edge qubit is shared with three other neighbors, a single
cell is effectively six qubits. The total plumbing piece volume in qubits is:

\begin{align}
    V &\approx 6 \left(\frac{5d}{4}\right)^3
\end{align}

We can decompose this into physical and time qubits as follows. Each cell
requires two layers of $3 n^2$ physical qubits, while the $n$ cells in
time results in a total of $6 n^3$.

Plumbing piece volume is measured in units of qubit-rounds. This takes into
account not only the physical resource overhead, but the additional time (in
rounds of error correction) necessary to implement larger distances. Focusing
on a single plumbing piece provides a measure of overhead independent of any
specific algorithm. This allows relative performance comparisons regardless of
future improvements to algorithms and circuit compression.

\section{Simulation}
\label{sec:sim}

Simulations were performed using an updated version of our \textit{Autotune}
software~\citep{Fowler:2012aa}. The logical error rate ($p_{L}$) of a given
error model is determined by a continuous simulation of an ever growing
topological cluster state. Each iteration of the simulation extends the problem
by $t_{check}$ rounds. The problem is then \textit{capped} by simulating two
more rounds, with all errors disabled, to ensure that all cells have been
completed. The detection events are then analysed to determine if a logical
error has occurred. Finally, the actions taken to cap the problem are reversed
in order to continue the simulation of the next $t_{check}$ rounds.

In more detail, each iteration begins with the simulation of
$t_{check}$ rounds of the topological cluster state. This is followed by all
errors being disabled and two more rounds of error correction being performed
to finalize any remaining cells. The value of $t_{check}$ is dependent on the
expected $p_{L}$, starting at $1$ when $p_{L}$ is expected to be high (large
$p_{comp}$ and $p_{loss}$) and up to $10^4$ when $p_{L}$ is expected to be low
(small $p_{comp}$ and $p_{loss}$). 

We then use the \textit{Blossom V}~\citep{Kolmogorov:2009aa} matching library
in order to match the detection events. Autotune was updated to make use of
this publicly available library to make it both more useful to other
researchers, and to make our results more readily reproducible. This comes
with the downside of being slower than our previous matching
library~\citep{Fowler:2012ab}.

The Autotune software generates weights for the matching problem by analyzing
the user provided arbitrary stochastic error model to determine where and when
every possible error is detected. Many different errors can lead to the same
pair of detection events, and the total probability of all such errors can be
converted to a useful weight by taking the negative log. These weights are
stored in a 3-D graph structure, so that they can be used throughout the
simulation. By merging the weights of the underlying lattice when qubits are
lost, the matching can accurately account for these errors. 

After matching we can determine if a logical error has occurred by observing 
if there has been a change in the number of errors along a boundary.  Finally,
the two perfect rounds of error correction are undone and allowing the
simulation of another $t_{check}$ rounds.

In order to manage the problem size and memory use, detection events and
associated data are removed after $t_{delete}$ rounds have occurred. This is
possible due to it becoming highly unlikely that a detection event sufficiently
far in the past needs to be rematched. The value of $t_{delete}$ was chosen
to be $5 d$. 

The simulations were performed in this manner as it most closely resembles the
operation of an actual quantum computer. They were designed so that it was
feasible to replace the input error model with one generated from
experiment~\citep{Fowler:2014aa}, and use experimental measurement results as
opposed to simulating them. 

\section{Results}
\label{sec:results}

For the model including initialization, two-qubit and measurement loss errors,
we dynamically generated graphs plotting the probability of logical error
($P_{L}$) against the probability of loss ($p_{loss}$). We generated these
graphs for three fixed amounts of computational error: $p_{comp} = 0$ to
determine the behavior of loss alone; and $p_{comp} \in \{10^{-3}, 10^{-4}\}$
to determine the behavior of loss at two current experimental targets for
computational error. Under this error model we observe a threshold loss error
rate of $\sim$$2$--$5\%$. 

\begin{figure}[p]
\resizebox{80mm}{!}{\includegraphics{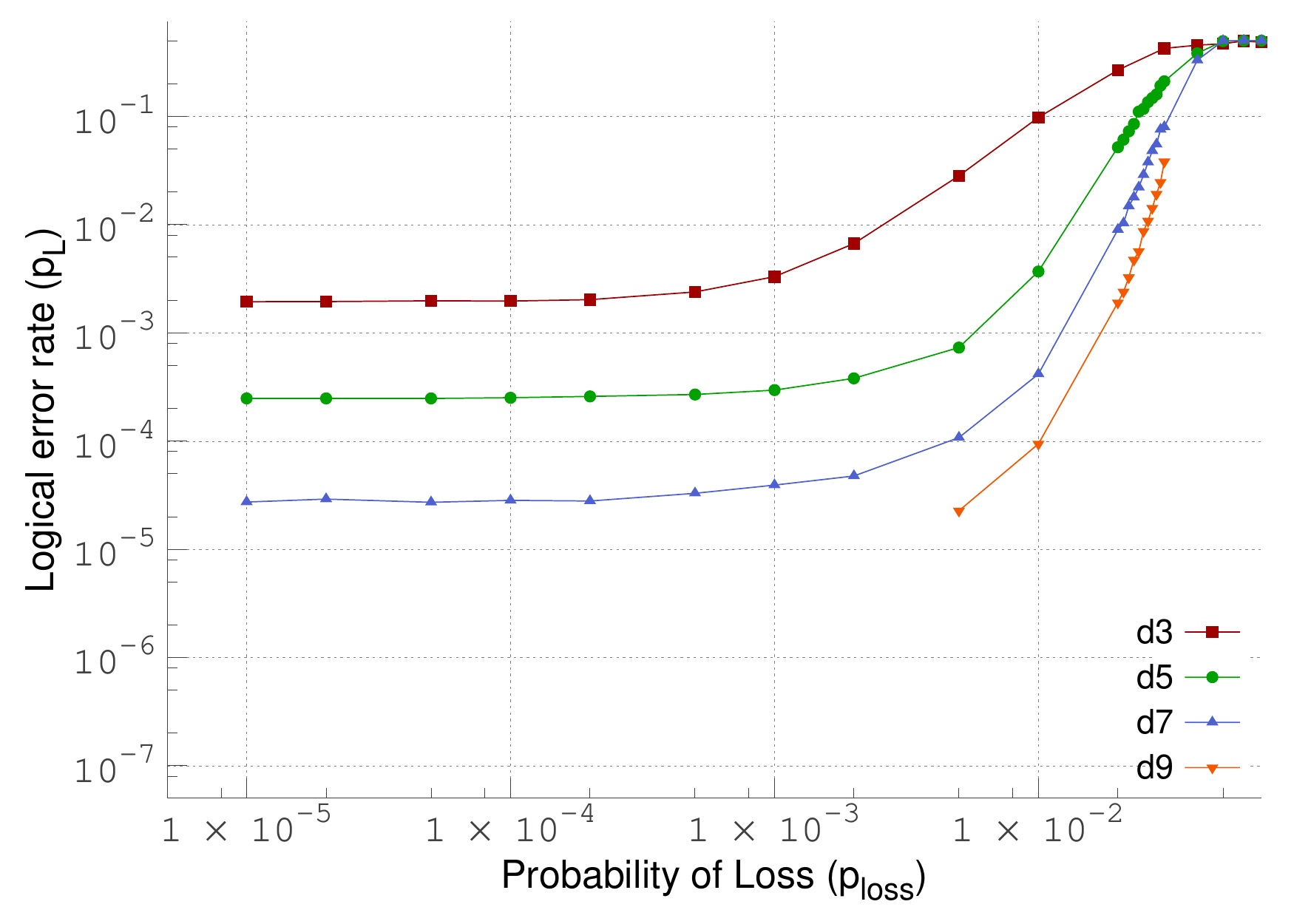}}
\caption{
    Logical error rate ($P_{L}$) vs.~Probability of loss ($p_{loss}$). Fixed
    computational error ($p_{comp} = 10^{-3}$) and no loss interaction error
    ($p_{lint} = 0$).
}\label{fig:graph_nolint_0001}
\end{figure}

With computational error of $0.1\%$ ($p_{comp} = 10^{-3}$)
(Fig.~\ref{fig:graph_nolint_0001}), the effect of loss is negligible when the
rate of loss is less than $0.01\%$ ($p_{loss} = 10^{-4}$). The
curves asymptote to the behavior obtained when only computational error
is considered ($p_{loss} = 0$), as expected~\citep{Fowler:2012aa}. 

\begin{figure}[p]
\resizebox{80mm}{!}{\includegraphics{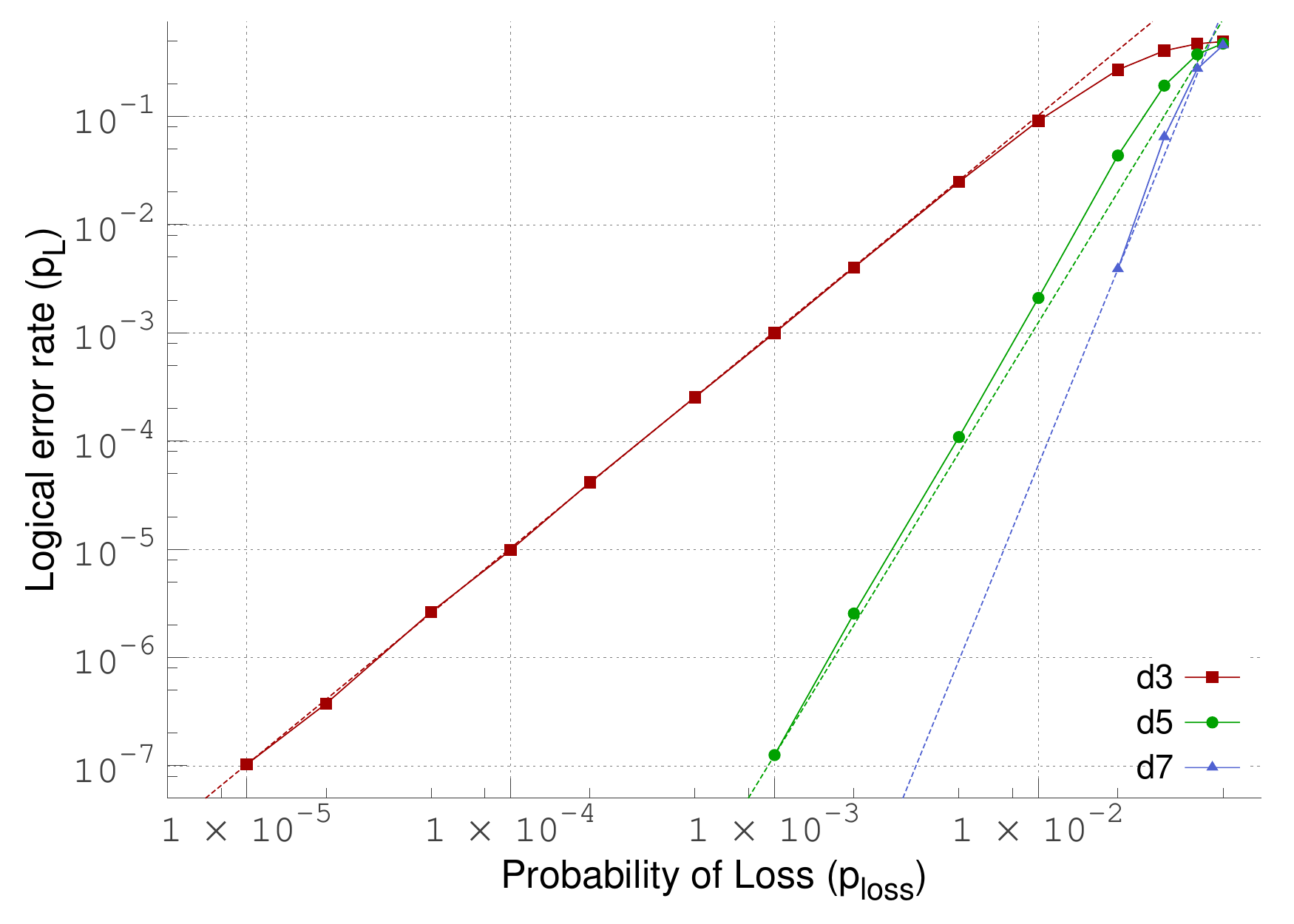}}
\caption{
    Logical error rate ($P_{L}$) vs.~Probability of loss ($p_{loss}$). Fixed
    computational error ($p_{comp} = 0$) and no loss interaction error
    ($p_{lint} = 0$). Dotted lines indicate asymptotic lines where logical
    errors predominantly caused by the minimum necessary ($\approx c
    p_{loss}^{d - 1}$).
}\label{fig:graph_nolint_0}
\end{figure}

We verify the correct behavior of our loss implementation by observing that
with no computational error ($p_{comp} = 0$) (Fig.~\ref{fig:graph_nolint_0}),
the curves asymptote toward functions defined by the minimum number of errors
that can lead to a logical failure, $d - 1$. With no computational error, the
only form of logical error is when a stabilizer has been merged to the
point of bordering both boundaries of the same type, then needs to be merged
with the boundary. That is, at sufficiently
low $p_{loss}$ the logical error rate can be approximated by:

\begin{align}
    P_{L} \approx c p_{loss}^{d - 1}
\end{align}

\begin{figure}[p]
\resizebox{80mm}{!}{\includegraphics{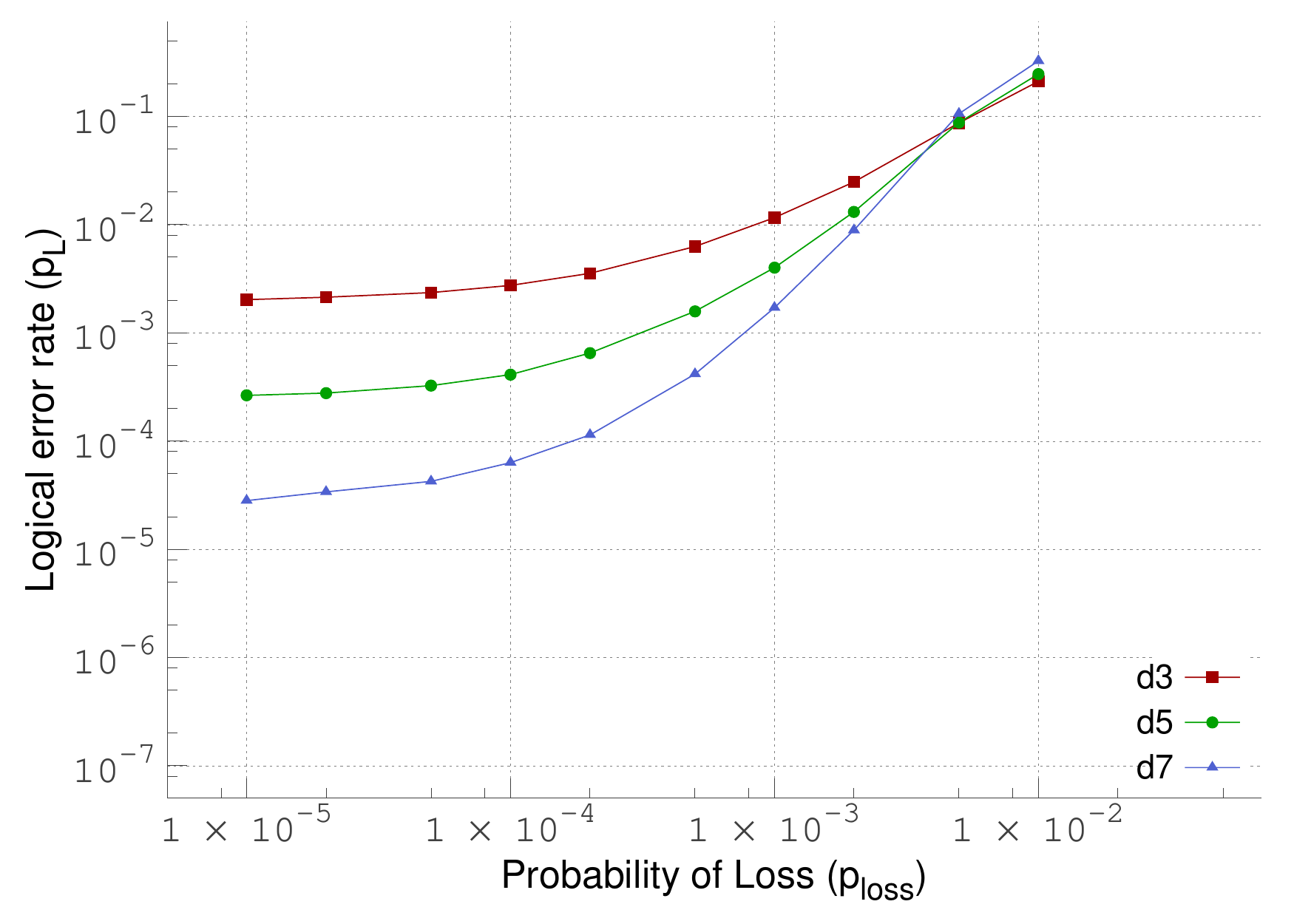}}
\caption{
    Logical error rate ($P_{L}$) vs.~Probability of loss ($p_{loss}$). Fixed
    computational error ($p_{comp} = 10^{-3}$) and 100\% chance of loss
    interaction error ($p_{lint} = 1$).
}\label{fig:graph_lint_0001}
\end{figure}

For the model where loss interaction errors are also included ($p_{lint} = 1$),
we observe an order of magnitude drop in the threshold error rate to
$\sim$$0.2$--$0.5\%$. With high computational error ($p_{comp} = 10^{-3}$)
(Fig.~\ref{fig:graph_lint_0001}) we observe that the impact of loss becomes
negligible when the rate of loss is an order of magnitude lower than the model
with no interaction errors ($p_{loss} = 10^{-5}$).

\begin{figure}[p]
\resizebox{80mm}{!}{\includegraphics{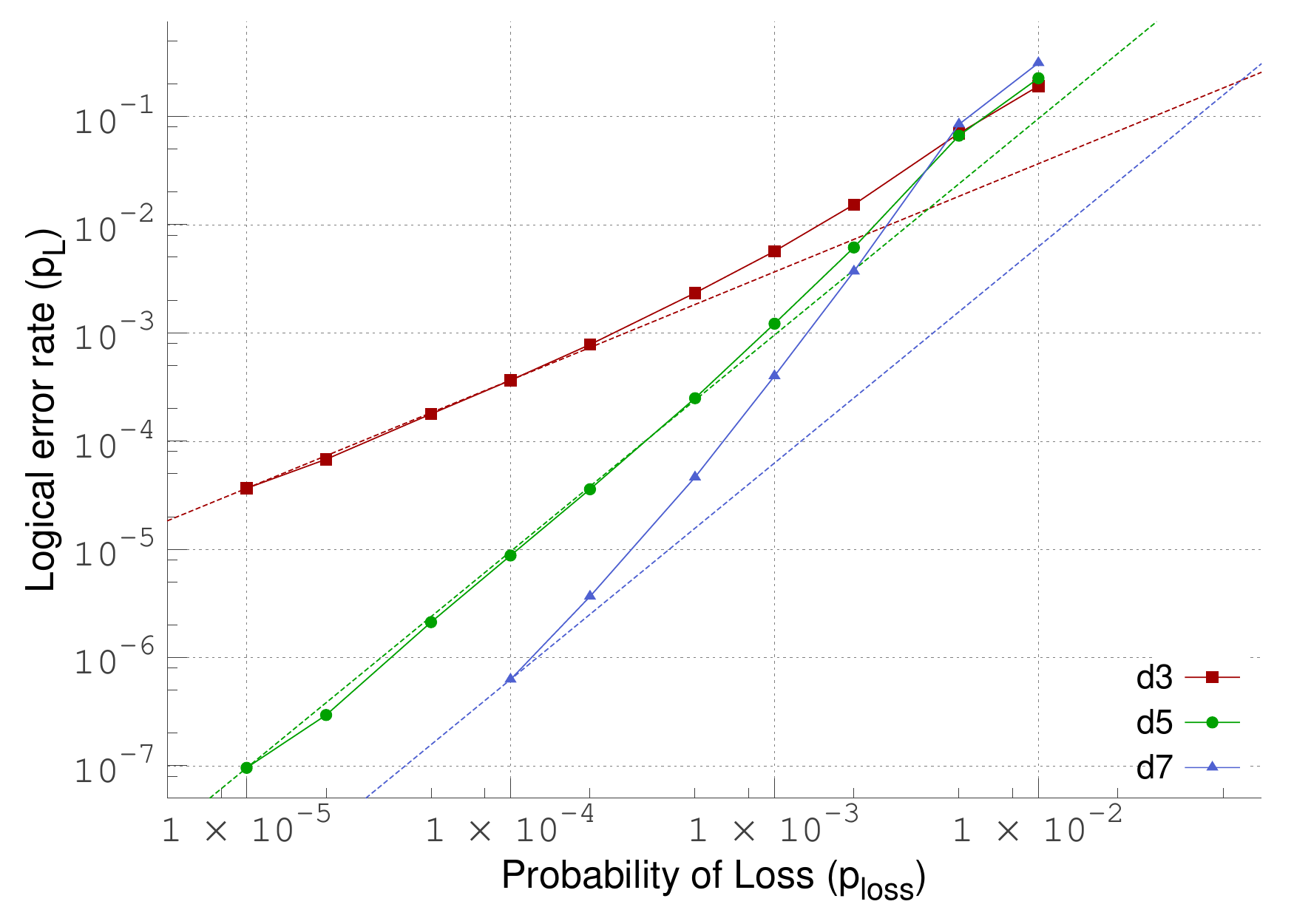}}
\caption{
    Logical error rate ($P_{L}$) vs.~Probability of loss ($p_{loss}$). Fixed
    computational error ($p_{comp} = 0$) and 100\% chance of loss interaction
    error ($p_{lint} = 1$). Dotted lines indicate asymptotic lines where
    logical errors predominantly caused by the minimum necessary ($\approx c
    p_{loss}^{\floor[\big]{\frac{d + 3}{4}}}$).
}\label{fig:graph_lint_0}
\end{figure}

A single loss event can form a chain of two computational errors when loss
interaction errors occur. This allows one loss error to act as two errors,
reducing the minimum number of errors required to cause a logical error to
$\floor[\big]{\frac{d + 3}{4}}$, adjusting the asymptotics as observed (Fig.~\ref{fig:graph_lint_0}).

\begin{figure}[p]
\resizebox{80mm}{!}{\includegraphics{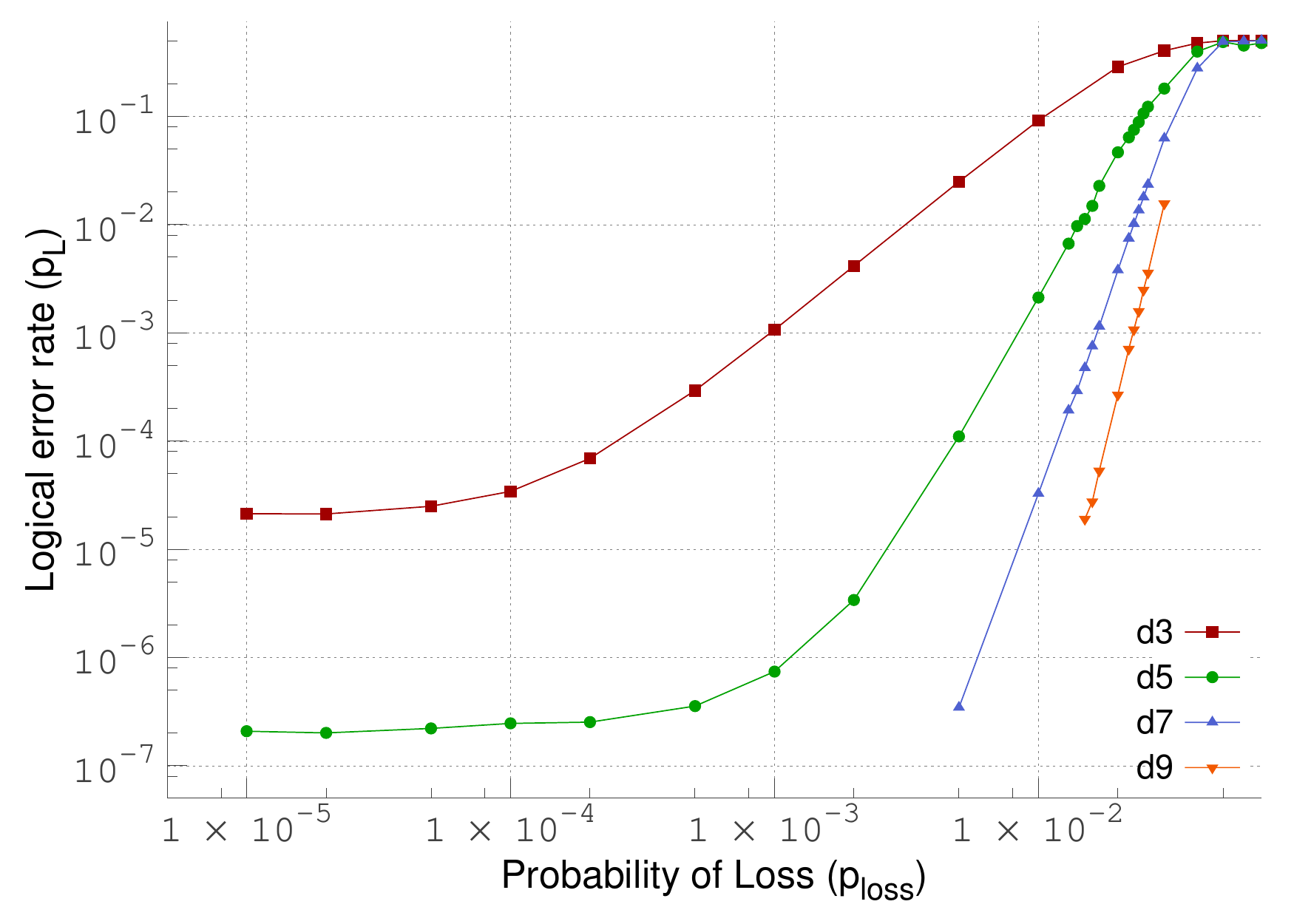}}
\caption{
    Logical error rate ($P_{L}$) vs.~Probability of loss ($p_{loss}$). Fixed
    computational error ($p_{comp} = 10^{-4}$) and no loss interaction error
    ($p_{lint} = 0$).
}\label{fig:graph_nolint_00001}
\end{figure}

\begin{figure}[p]
\resizebox{80mm}{!}{\includegraphics{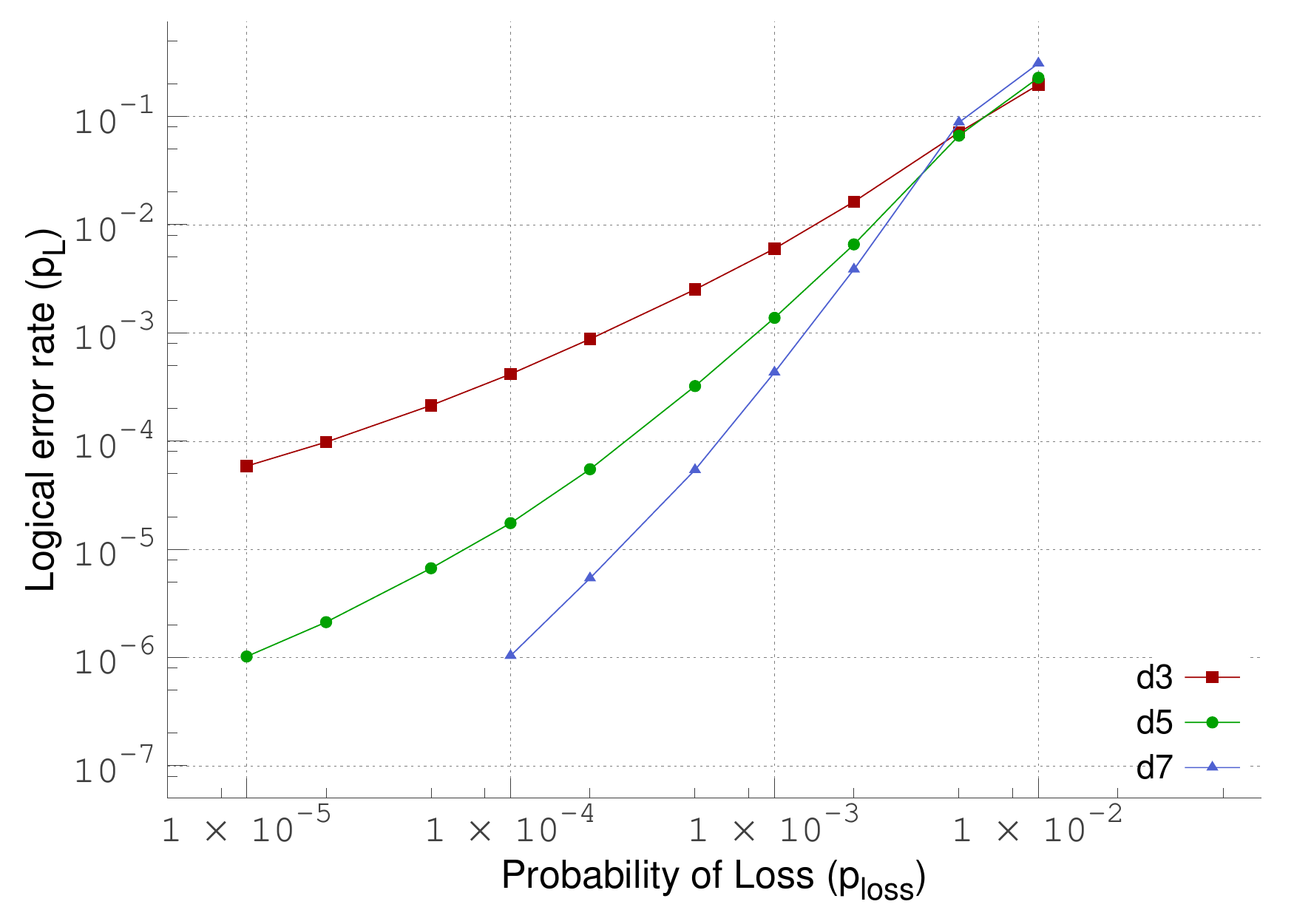}}
\caption{
    Logical error rate ($P_{L}$) vs.~Probability of loss ($p_{loss}$). Fixed
    computational error ($p_{comp} = 10^{-4}$) and 100\% chance of loss
    interaction error ($p_{lint} = 1$).
}\label{fig:graph_lint_00001}
\end{figure}

\begin{table}[p]
    \centering
    \renewcommand{\arraystretch}{1.1}
    \begin{tabular}{| >{$}c<{$} | >{$}c<{$} | >{$}c<{$} | >{$}c<{$} | >{$}c<{$} | >{$}c<{$} | >{$}c<{$} |}
    \hline
    \textbf{Overhead} & \mathbf{p_{Loss}} & \mathbf{d} & \mathbf{V} & \mathbf{q_{phys}} \\
    \hline
    1 \times & \text{None} & 31 & 3.2 \times 10^{5} & 8.6 \times 10^{3} \\
    \hline
    1 \times & 2.0 \times 10^{-3} & 31 & 3.2 \times 10^{5} & 8.6 \times 10^{3} \\
    2 \times & 1.0 \times 10^{-2} & 37 & 5.6 \times 10^{5} & 1.2 \times 10^{4} \\
    5 \times & 2.5 \times 10^{-2} & 51 & 1.5 \times 10^{6} & 2.4 \times 10^{4} \\
    10 \times & 2.9 \times 10^{-2} & 67 & 3.4 \times 10^{6} & 4.1 \times 10^{4} \\    
    \text{-} & 5.0 \times 10^{-2} & \text{-} & \text{-} & \text{-} \\
    \hline
\end{tabular}
\caption{%
    Table of volume and physical qubit overheads at supplied rates of loss. 
    Computational error rate: $p_{comp} = 10^{-3}$. 
    Loss interaction error rate: $p_{lint} = 0$. 
    Target logical error rate: $P_L = 10^{-15}.$  
    Higher distances extrapolated from the behavior of distances $5, 7, 9$.
}
\label{tbl:nolint:0.001}
\end{table}

\begin{table}[p]
    \centering
    \renewcommand{\arraystretch}{1.1}
    \begin{tabular}{| >{$}c<{$} | >{$}c<{$} | >{$}c<{$} | >{$}c<{$} | >{$}c<{$} | >{$}c<{$} | >{$}c<{$} |}
    \hline
    \textbf{Overhead} & \mathbf{p_{loss}} & \mathbf{d} & \mathbf{V} & \mathbf{q_{phys}} \\
    \hline
    1 \times & \text{None} & 15 & 3.4 \times 10^{4} & 1.9 \times 10^{3} \\   
    \hline
    1 \times & 5.0 \times 10^{-3} & 15 & 3.4 \times 10^{4} & 1.9 \times 10^{3} \\
    2 \times & 1.0 \times 10^{-2} & 19 & 7.1 \times 10^{4}  & 3.1 \times 10^{3} \\
    8 \times & 2.0 \times 10^{-2} & 29 & 2.6 \times 10^{5}  & 7.5 \times 10^{3} \\
    10 \times & 2.2 \times 10^{-2} & 31 & 3.2 \times 10^{5}  & 8.6 \times 10^{3} \\
    \text{-} & 5.0 \times 10^{-2} &  \text{-} & \text{-} & \text{-} \\
    \hline
\end{tabular}
\caption{%
    Table of volume and physical qubit overheads at supplied rates of loss. 
    Computational error rate: $p_{comp} = 10^{-4}$. 
    Loss interaction error rate: $p_{lint} = 0$. 
    Target logical error rate: $P_L = 10^{-15}.$  
    Higher distances extrapolated from the behavior of distances $5, 7, 9$.
}
\label{tbl:nolint:0.0001}
\end{table}

\begin{table}[p]
    \centering
    \renewcommand{\arraystretch}{1.1}
    \begin{tabular}{| >{$}c<{$} | >{$}c<{$} | >{$}c<{$} | >{$}c<{$} | >{$}c<{$} | >{$}c<{$} | >{$}c<{$} |}
    \hline
    \textbf{Overhead} & \mathbf{p_{Loss}} & \mathbf{d} & \mathbf{V} & \mathbf{q_{phys}} \\
    \hline
    1 \times & \text{None} & 31 & 3.2 \times 10^{5} & 8.6 \times 10^{3} \\
    \hline
    1 \times & 1.0 \times 10^{-4} & 33 & 3.9 \times 10^{5} & 9.7 \times 10^{3} \\
    2 \times & 2.0 \times 10^{-4} & 37 & 5.6 \times 10^{5} & 1.2 \times 10^{4} \\
    3 \times & 5.0 \times 10^{-4} & 45 & 1.0 \times 10^{6} & 1.8 \times 10^{4} \\
    10 \times & 1.0 \times 10^{-3} & 63 & 2.8 \times 10^{6} & 3.6 \times 10^{4} \\
    \text{-} & 5.0 \times 10^{-3} & \text{-} & \text{-} & \text{-} \\
    \hline
\end{tabular}
\caption{%
    Table of volume and physical qubit overheads at supplied rates of loss. 
    Computational error rate: $p_{comp} = 10^{-3}$. 
    Loss interaction error rate: $p_{lint} = 1$. 
    Target logical error rate: $P_L = 10^{-15}.$  
    Higher distances extrapolated from the behavior of distances $3, 5, 7$.
}
\label{tbl:lint:0.001}
\end{table}

\begin{table}[p]
    \centering
    \renewcommand{\arraystretch}{1.1}
    \begin{tabular}{| >{$}c<{$} | >{$}c<{$} | >{$}c<{$} | >{$}c<{$} | >{$}c<{$} | >{$}c<{$} | >{$}c<{$} |}
    \hline
    \textbf{Overhead} & \mathbf{p_{Loss}} & \mathbf{d} & \mathbf{V} & \mathbf{q_{phys}} \\
    \hline
    1 \times & \text{None} & 15 & 3.4 \times 10^{4} & 1.9 \times 10^{3} \\
    \hline
    1 \times & 1.0 \times 10^{-5} & 15 & 3.4 \times 10^{4} & 1.9 \times 10^{3} \\
    2 \times & 5.0 \times 10^{-5} & 19 & 7.1 \times 10^{4} & 3.1 \times 10^{3} \\
    5 \times & 2.0 \times 10^{-4} & 25 & 1.7 \times 10^{5} & 5.5 \times 10^{3} \\
    10 \times & 5.0 \times 10^{-4} & 33 & 3.9 \times 10^{5} & 9.7 \times 10^{3} \\
    \text{-} & 5.0 \times 10^{-3} & \text{-} & \text{-} & \text{-} \\
    \hline
\end{tabular}
\caption{%
    Table of volume and physical qubit overheads at supplied rates of loss. 
    Computational error rate: $p_{comp} = 10^{-4}$. 
    Loss interaction error rate: $p_{lint} = 1$. 
    Target logical error rate: $P_L = 10^{-15}.$  
    Higher distances extrapolated from the behavior of distances $3, 5, 7$.
}
\label{tbl:lint:0.0001}
\end{table}

Four tables of results were calculated, consisting of each paired combination
of $p_{lint} \in \{0,1\}$ and $p_{comp} \in \{10^{-3}, 10^{-4}\}$. Overhead is
calculated using the number of qubits necessary to create a plumbing piece
achieving a given logical error rate ($P_L = 10^{-15}$ in all instances) in
comparison to the baseline with no loss ($p_{loss} = 0$). The values used for
the baseline are taken or extrapolated from the data presented
in~\citep{Fowler:2012aa}.

Data was collected for distances $3$, $5$, $7$ and $9$, higher distances were
extrapolated from the average error suppression ratio of the two highest
available distances. At low $d$, the ratios of error suppression for points
near to the threshold are not expected to be constant. As a result,
extrapolation at these points will underestimate the overhead. Sufficiently far
from the threshold, or at large $d$, this error suppression ratio is expected
to be constant~\citep{Fowler:2012ah}. Direct simulation and extrapolation have
been compared in detail for the case of the surface code~\citep{Fowler:2013ad}
with a high level of agreement when error models are not asymmetric, the case
studied here.

We now provide an example of this extrapolation for clarity. To extrapolate to
any distance we take the value for highest distance and divide it by the ratio
of the two highest distance points the appropriate number of times. Let $a$
and $b$ be the second highest and highest distance points respectively. $d$ is
the desired distance, and $d_{b}$ is the distance of the highest point.

\[
    p_L \approx \frac{b}{\left(\frac{a}{b}\right)^{\left\lfloor\frac{d - d_{b}}{2}\right\rfloor}}
\]

With loss interaction errors enabled ($p_{lint} = 1$), a computational error
rate of $0.1\%$ ($p_{comp} = 0.001$) and loss rate of $0.01\%$ ($p_{loss} =
10^{-4}$) the logical error rates for distances $5$ and $7$ are $a = 4.1 \times
10^{-4}$ and $b = 6.3 \times 10^{-5}$ respectively (Fig
~\ref{fig:graph_lint_0001}). The ratio between these two points is therefore
$\sim$$6.51$ and $d_{b} = 7$. At distance $33$, this gives us a value for $p_L
\approx 1.7 \times 10^{-15}$ which is below our target threshold of $p_L =
10^{-15}$.

Without loss interaction errors ($p_{lint} = 0$), we find for $p_{comp} \in
\{10^{-3}, 10^{-4}\}$
(Tables~\ref{tbl:nolint:0.001} \& \ref{tbl:nolint:0.0001}) that values of
$p_{loss} \ge 5\%$ are clearly above threshold, with the actual threshold
expected to be between $3$--$4\%$. It is observed that $\sim$$2.5\%$ loss is
tolerable with $\le 10 \times$ the qubit volume required to implement the same
algorithm as an identical quantum computer with no loss. For values of
$p_{loss}$ above this point, the overhead rapidly increases. 

The topological cluster state has the highest threshold of any code that
requires only a 2D lattice of qubits with nearest-neighbor interactions, while
also having a process for handling qubit loss. Despite requiring the fewest
qubits to implement a logical qubit of any code under these practical hardware
restrictions, the number of qubits necessary is high. Given a single logical
qubit needs $\sim$$3 \times 10^{4}$ qubit-rounds without loss present, we
choose to consider a loss rate that more than doubles the baseline overhead to
be undesirable.

An architecture with loss during initialization, two-qubit gates, and
measurement should therefore highly preferably have a loss rate less than $1\%$
in order to keep the overhead under control. In order for loss to have a
negligible penalty -- such that
there is no overhead compared to when there is no loss at all -- a loss rate
less than $0.5\%$ is necessary.

Due to the larger effective impact of loss when loss interaction errors can
occur ($p_{lint} = 1$), and the consequent reduction of the gate threshold error
rate to not far above $10^{-3}$, there are more substantial differences in the
overheads between $p_{comp} = 10^{-3}$ (Table~\ref{tbl:lint:0.001}) and
$p_{comp} = 10^{-4}$ (Table~\ref{tbl:lint:0.0001}). When loss interaction
errors occur, values of $p_{loss} \ge 0.5\%$ are clearly above threshold, with
the actual threshold expected to be between $0.3$--$0.4\%$. The maximum amount of loss
tolerable with $\le 10 \times$ overhead is approximately $0.05$--$0.1\%$, with
$0.001$--$0.01\%$ necessary to have no overhead.

The rate of loss tolerable while maintaining modest ($\le2\times$) overhead
is as low as $0.005\%$ with $0.1\%$ computational error ($p_{comp} = 10^{-3}$),
considerably lower than when there are no loss interaction errors. 

Given our focus on practical overhead, we have omitted techniques that would
allow us to tolerate more loss. Techniques exist, particularly for linear
optical architectures, which allow substantial improvement to the acceptable
amount of loss at the cost of introducing substantial
overhead~\citep{Nickerson:2014aa,Varnava:2008aa,Li:2012aa}.

\section{Discussion}
\label{sec:discussion}

We have shown that when including loss during initialization and two-qubit
gates, the amount of loss tolerable is much smaller than when only considering
loss before measurement. We find that even in the best case scenario ($10^{-4}$
computational error and no errors caused when interacting with lost qubits)
only $1\%$ loss can be tolerated if we allow the use of twice the qubit volume
as would be required by a lossless quantum computer. The quickly increasing
overhead beyond $1\%$ (threshold $3$--$4\%$) rapidly becomes highly
impractical.  These results, combined with the fact our results are likely to
underestimate the overhead, lead us to conclude that even in the best case, no
more than $1\%$ loss can be tolerated while maintaining practical overheads. 

We have also demonstrated the importance of considering loss interaction
errors. The creation of additional errors on neighboring qubits when they
attempt to interact with a lost qubit reduces the amount of loss that is
practically tolerable by at least an order of magnitude, down to $0.1\%$ for a
$10$$\times$ overhead. This is due to a single loss error potentially causing a
chain of two computational errors.

We have provided the first analysis of the topological cluster state for its
tolerance to loss under these more detailed error models, as well as providing
overhead scaling for two current experimental targets, $p_{comp} \in \{10^{-3},
10^{-4}\}$. In doing so, we have shown that by ignoring initialization and
two-qubit loss, previous work has overestimated the amount of loss tolerable.
Our work provides concrete experimental loss targets for a range of resultant
overheads, and highlights the importance of reducing or eliminating loss
interaction errors.

\section{Acknowledgments}
\label{ack}

This research was funded by the Australian Federal Government. This research
was supported in part by the Australian Research Council Centre of Excellence
for Quantum Computation and Communication Technology (project number
CE110001027). AGF was funded by the US Office of the Director of National
Intelligence (ODNI), Intelligence Advanced Research Projects Activity (IARPA),
through the US Army Research Office grant No. W911NF-10-1-0334. All statements
of fact, opinion or conclusions contained herein are those of the authors and
should not be construed as representing the official views or policies of
IARPA, the ODNI, or the US Government. 

\newpage
\bibliography{loss}

\appendix

\newpage

\section{Derivation of Cell State}
\label{app:cell}

\begin{figure}
\begin{center}
\resizebox{89mm}{!}{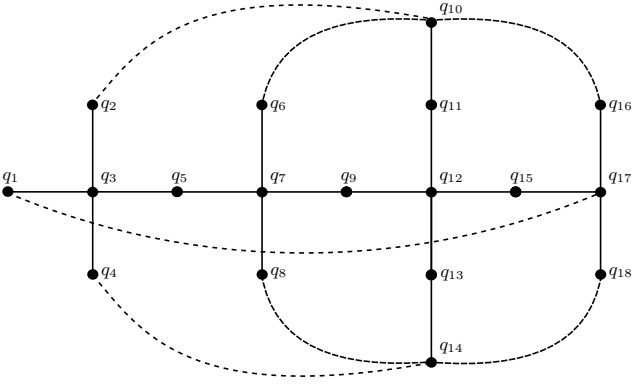}
\end{center}
\caption{
    A 2D representation of the 18 qubit cluster state of a single 3D
    topological cluster state cell (Fig.~\ref{fig:tcscell}). Black dots are the
    qubits $q_1$ to $q_18$, initialized in the $\ket{+}$ state. Black lines
    indicate $C_Z$ gates between the two connected qubits. Dotted lines are a
    visual guide for $C_Z$ gates that are not nearest neighbors in the 2D
    flattening.
}\label{fig:cellstabilizer}
\end{figure}

In Table \ref{tbl:cellstabilizers} we can see a set of stabilizers for the 3D
topological cluster state cell (Fig.~\ref{fig:cellstabilizer}). This is just one possible generating set for
the given cluster state. It was obtained according to the procedure outlined
in Section \ref{sec:cs}.

\begin{center}
    \begin{table}[b]
    \renewcommand{\arraystretch}{1.0}
    \begin{tabular}{ >{$}l<{$} | >{$}c<{$} >{$}c<{$} >{$}c<{$} >{$}c<{$} >{$}c<{$} >{$}c<{$} >{$}c<{$} >{$}c<{$} >{$}c<{$} >{$}c<{$} >{$}c<{$} >{$}c<{$} >{$}c<{$} >{$}c<{$} >{$}c<{$} >{$}c<{$} >{$}c<{$} >{$}c<{$} }
           A_1 & X & I & Z & I & I & I & I & I & I & I & I & I & I & I & I & I & Z & I \\
           A_2 & I & X & Z & I & I & I & I & I & I & Z & I & I & I & I & I & I & I & I \\
           A_3 & Z & Z & X & Z & Z & I & I & I & I & I & I & I & I & I & I & I & I & I \\
           A_4 & I & I & Z & X & I & I & I & I & I & I & I & I & I & Z & I & I & I & I \\
           A_5 & I & I & Z & I & X & I & Z & I & I & I & I & I & I & I & I & I & I & I \\
           A_6 & I & I & I & I & I & X & Z & I & I & Z & I & I & I & I & I & I & I & I \\
           A_7 & I & I & I & I & Z & Z & X & Z & Z & I & I & I & I & I & I & I & I & I \\
           A_8 & I & I & I & I & I & I & Z & X & I & I & I & I & I & Z & I & I & I & I \\
           A_9 & I & I & I & I & I & I & Z & I & X & I & I & Z & I & I & I & I & I & I \\
        A_{10} & I & Z & I & I & I & Z & I & I & I & X & Z & I & I & I & I & Z & I & I \\
        A_{11} & I & I & I & I & I & I & I & I & I & Z & X & Z & I & I & I & I & I & I \\
        A_{12} & I & I & I & I & I & I & I & I & Z & I & Z & X & Z & I & Z & I & I & I \\
        A_{13} & I & I & I & I & I & I & I & I & I & I & I & Z & X & Z & I & I & I & I \\
        A_{14} & I & I & I & Z & I & I & I & Z & I & I & I & I & Z & X & I & I & I & Z \\
        A_{15} & I & I & I & I & I & I & I & I & I & I & I & Z & I & I & X & I & Z & I \\
        A_{16} & I & I & I & I & I & I & I & I & I & Z & I & I & I & I & I & X & Z & I \\
        A_{17} & Z & I & I & I & I & I & I & I & I & I & I & I & I & I & Z & Z & X & Z \\
        A_{18} & I & I & I & I & I & I & I & I & I & I & I & I & I & Z & I & I & Z & X
    \end{tabular}
        \caption{
            A set of stabilizers for a single 3D topological cluster
            state cube, containing one for each of the 18 qubits. After the
            application of the $C_Z$ gates, the cell will be simultaneously in the $+1$
            eigenstate of each of these stabilizers. 
        }
        \label{tbl:cellstabilizers}
    \end{table}
\end{center}

If we select the stabilizers we have created based on each of the face qubits,
$A_3$, $A_7$, $A_{10}$, $A_{12}$, $A_{14}$ and $A_{17}$ (Table
\ref{tbl:cellsubset}). It can be seen, that each column of the stabilizer
generating set contains either a single $\hat{X}$ operator, or two $\hat{Z}$ operators and
$\hat{I}$ on the remainder.

\begin{center}
    \begin{table}
    \renewcommand{\arraystretch}{1.1}
    \begin{tabular}{ >{$}l<{$} | >{$}c<{$} >{$}c<{$} >{$}c<{$} >{$}c<{$} >{$}c<{$} >{$}c<{$} >{$}c<{$} >{$}c<{$} >{$}c<{$} >{$}c<{$} >{$}c<{$} >{$}c<{$} >{$}c<{$} >{$}c<{$} >{$}c<{$} >{$}c<{$} >{$}c<{$} >{$}c<{$} }
           A_3 & Z & Z & X & Z & Z & I & I & I & I & I & I & I & I & I & I & I & I & I \\
           A_7 & I & I & I & I & Z & Z & X & Z & Z & I & I & I & I & I & I & I & I & I \\
        A_{10} & I & Z & I & I & I & Z & I & I & I & X & Z & I & I & I & I & Z & I & I \\
        A_{12} & I & I & I & I & I & I & I & I & Z & I & Z & X & Z & I & Z & I & I & I \\
        A_{14} & I & I & I & Z & I & I & I & Z & I & I & I & I & Z & X & I & I & I & Z \\
        A_{17} & Z & I & I & I & I & I & I & I & I & I & I & I & I & I & Z & Z & X & Z
    \end{tabular}
        \caption{
            A subset of the stabilizers for a single 3D topological cluster
            state cube. Specifically, the subset that were stabilizing each of
            the center face qubits of the 3D topological cluster state cell.
        }
        \label{tbl:cellsubset}
    \end{table}
\end{center}

If we then take the matrix product of these stabilizers we can form a new
stabilizer, $A$ (Table \ref{tbl:cellstabilizer}). Given that $(A_{1} \otimes
B_{2}) \cdot (C_{1} \otimes D_{2}) = (A_{1} \cdot C{_1}) \otimes (B_{2} \cdot
D_{2})$, the $\hat{Z}$ operators will all cancel (as the inverse of $\hat{Z}$
is $\hat{Z}$) and we will be left with a new stabilizer with $\hat{X}$
operators on the face qubits ($q_3$, $q_7$, $q_{10}$, $q_{12}$, $q_{14}$ and
$q_{17}$) and $I$ on the others.

\begin{center}
    \begin{table}
    \renewcommand{\arraystretch}{1.1}
    \begin{tabular}{ >{$}l<{$} | >{$}c<{$} >{$}c<{$} >{$}c<{$} >{$}c<{$} >{$}c<{$} >{$}c<{$} >{$}c<{$} >{$}c<{$} >{$}c<{$} >{$}c<{$} >{$}c<{$} >{$}c<{$} >{$}c<{$} >{$}c<{$} >{$}c<{$} >{$}c<{$} >{$}c<{$} >{$}c<{$} }
           A & I & I & X & I & I & I & X & I & I & X & I & X & I & X & I & I & X & I
    \end{tabular}
        \caption{
            A new stabilizer $A$ that is the matrix product of $A_3$, $A_7$,
            $A_{10}$, $A_{12}$, $A_{14}$ and $A_{17}$. This can replace any of
            the 6 stabilizers to form a new valid set of stabilizers. 
        }
        \label{tbl:cellstabilizer}
    \end{table}
\end{center}

If we measure the new stabilizer $A$ (by measuring the face qubits in the
$\hat{X}$ basis), in the absence of errors, each qubit will be in either the
$+1$ or $-1$ eigenstate of $\hat{X}$. On the whole however, the result will be
the $+1$ eigenstate of the stabilizer $A$. Therefore, if you take the product
of each of the measurement results, the result (in the absence of error) will
be $+1$.  It is important to note that the stabilizer $A$ does not force each
qubit to be in the $+1$ eigenstate of $\hat{X}$, only that the measurement
product of all six face qubits be $+1$.

The reason we chose the stabilizers $A_3$, $A_7$, $A_{10}$, $A_{12}$, $A_{14}$
and $A_{17}$ is because we can only measure in a single basis simultaneously. 
Therefore, we want a stabilizer that consists of only $\hat{X}$ or only $\hat{Z}$
operators. The stabilizers chosen are the easiest way to obtain this given the
method of generating the initial set of stabilizers for the 3D topological
cluster cell state.

\end{document}

%% file: facestabilizer.pdf_tex
\begingroup%
  \makeatletter%
  \providecommand\color[2][]{%
    \errmessage{(Inkscape) Color is used for the text in Inkscape, but the package 'color.sty' is not loaded}%
    \renewcommand\color[2][]{}%
  }%
  \providecommand\transparent[1]{%
    \errmessage{(Inkscape) Transparency is used (non-zero) for the text in Inkscape, but the package 'transparent.sty' is not loaded}%
    \renewcommand\transparent[1]{}%
  }%
  \providecommand\rotatebox[2]{#2}%
  \ifx\svgwidth\undefined%
    \setlength{\unitlength}{102.97189629bp}%
    \ifx\svgscale\undefined%
      \relax%
    \else%
      \setlength{\unitlength}{\unitlength * \real{\svgscale}}%
    \fi%
  \else%
    \setlength{\unitlength}{\svgwidth}%
  \fi%
  \global\let\svgwidth\undefined%
  \global\let\svgscale\undefined%
  \makeatother%
  \begin{picture}(1,0.87710451)%
    \put(0,0){\includegraphics[width=\unitlength]{facestabilizer.pdf}}%
    \put(0.47512848,0.82988289){\color[rgb]{0,0,0}\makebox(0,0)[lb]{\smash{$q_1$}}}%
    \put(0.47651353,0.02365542){\color[rgb]{0,0,0}\makebox(0,0)[lb]{\smash{$q_5$}}}%
    \put(0.81593886,0.4808053){\color[rgb]{0,0,0}\makebox(0,0)[lb]{\smash{$q_4$}}}%
    \put(0.47455151,0.4808053){\color[rgb]{0,0,0}\makebox(0,0)[lb]{\smash{$q_3$}}}%
    \put(0.0017103,0.4808053){\color[rgb]{0,0,0}\makebox(0,0)[lb]{\smash{$q_2$}}}%
  \end{picture}%
\endgroup%

%% file: cellstabilizer.pdf_tex
\begingroup%
  \makeatletter%
  \providecommand\color[2][]{%
    \errmessage{(Inkscape) Color is used for the text in Inkscape, but the package 'color.sty' is not loaded}%
    \renewcommand\color[2][]{}%
  }%
  \providecommand\transparent[1]{%
    \errmessage{(Inkscape) Transparency is used (non-zero) for the text in Inkscape, but the package 'transparent.sty' is not loaded}%
    \renewcommand\transparent[1]{}%
  }%
  \providecommand\rotatebox[2]{#2}%
  \ifx\svgwidth\undefined%
    \setlength{\unitlength}{332.36268609bp}%
    \ifx\svgscale\undefined%
      \relax%
    \else%
      \setlength{\unitlength}{\unitlength * \real{\svgscale}}%
    \fi%
  \else%
    \setlength{\unitlength}{\svgwidth}%
  \fi%
  \global\let\svgwidth\undefined%
  \global\let\svgscale\undefined%
  \makeatother%
  \begin{picture}(1,0.56235578)%
    \put(0,0){\includegraphics[width=\unitlength]{cellstabilizer.pdf}}%
    \put(0.14720329,0.40242183){\color[rgb]{0,0,0}\makebox(0,0)[lb]{\smash{$q_2$}}}%
    \put(0.1476324,0.15263816){\color[rgb]{0,0,0}\makebox(0,0)[lb]{\smash{$q_4$}}}%
    \put(0.25279243,0.29427136){\color[rgb]{0,0,0}\makebox(0,0)[lb]{\smash{$q_5$}}}%
    \put(0.14702453,0.29427136){\color[rgb]{0,0,0}\makebox(0,0)[lb]{\smash{$q_3$}}}%
    \put(0.00052988,0.29427136){\color[rgb]{0,0,0}\makebox(0,0)[lb]{\smash{$q_1$}}}%
    \put(0.40015542,0.40228056){\color[rgb]{0,0,0}\makebox(0,0)[lb]{\smash{$q_6$}}}%
    \put(0.40058455,0.15249692){\color[rgb]{0,0,0}\makebox(0,0)[lb]{\smash{$q_8$}}}%
    \put(0.50574458,0.29413009){\color[rgb]{0,0,0}\makebox(0,0)[lb]{\smash{$q_9$}}}%
    \put(0.39997671,0.29413009){\color[rgb]{0,0,0}\makebox(0,0)[lb]{\smash{$q_7$}}}%
    \put(0.65302738,0.40228057){\color[rgb]{0,0,0}\makebox(0,0)[lb]{\smash{$q_{11}$}}}%
    \put(0.65431617,0.14991797){\color[rgb]{0,0,0}\makebox(0,0)[lb]{\smash{$q_{13}$}}}%
    \put(0.75861654,0.29413009){\color[rgb]{0,0,0}\makebox(0,0)[lb]{\smash{$q_{15}$}}}%
    \put(0.65284866,0.29413009){\color[rgb]{0,0,0}\makebox(0,0)[lb]{\smash{$q_{12}$}}}%
    \put(0.65284873,0.54772568){\color[rgb]{0,0,0}\makebox(0,0)[lb]{\smash{$q_{10}$}}}%
    \put(0.65284866,0.03967484){\color[rgb]{0,0,0}\makebox(0,0)[lb]{\smash{$q_{14}$}}}%
    \put(0.90576333,0.40228056){\color[rgb]{0,0,0}\makebox(0,0)[lb]{\smash{$q_{16}$}}}%
    \put(0.90619246,0.15249692){\color[rgb]{0,0,0}\makebox(0,0)[lb]{\smash{$q_{18}$}}}%
    \put(0.90558461,0.2941301){\color[rgb]{0,0,0}\makebox(0,0)[lb]{\smash{$q_{17}$}}}%
  \end{picture}%
\endgroup%